\documentclass[structabstract]{aa}  
\usepackage{graphicx}
\usepackage{txfonts}
\usepackage{longtable}
\usepackage{natbib}

\bibpunct{(}{)}{;}{a}{}{,}

\newcommand{\tastar}{\hbox {T$_{\rm a}^*$ }}

\newcommand{\twco}{{\hbox {\ensuremath{\mathrm{^{12}CO}} }}}
\newcommand{\ceo}{{\hbox {\ensuremath{\mathrm{C^{18}O}} }}}
\newcommand{\ceop}{{\hbox {\ensuremath{\mathrm{C^{18}O}}}}}

\newcommand{\htcopl}{{\hbox {\ensuremath{\mathrm{H^{13}CO^+}} }}}

\newcommand{\hcopl}{{\hbox {\ensuremath{\mathrm{HCO^+}}}\ }}

\newcommand{\kmps}{\ensuremath{\mathrm{km\,s^{-1}}}}
\newcommand{\Msun}{\ensuremath{\mathrm{M}_\odot}}

\newcommand{\Lsun}{\ensuremath{\mathrm{L}_\odot}}
\newcommand{\kkms}{\ensuremath{\mathrm{K\,km\,s^{-1}}}}

\newcommand{\Tmb}{\ensuremath{\mathrm{T}_\mathrm{MB}}}
\newcommand{\mum}{\mu {\rm m}}
\def\Ks{\hbox{$K$s}}
\def\K{\hbox{$K$}}
\def\Js{\hbox{$J$s}}
\def\J{\hbox{$J$}}
\def\H{\hbox{$H$}}
\newcommand{\umag}{{$^{m}$}}
 
\def\jhks{\hbox{$J\!H\!K$s}}              % JHKs system
\newcommand{\ceomax }{{\hbox {\ensuremath{\mathrm{C^{18}O}}\_maximum}}}
\newcommand{\ceoSE }{{\hbox {\ensuremath{\mathrm{C^{18}O}}\_SE}}}
\newcommand{\ceoNW }{{\hbox {\ensuremath{\mathrm{C^{18}O}}\_NW}}}
\newcommand{\Htwo}{{\hbox {\ensuremath{\mathrm{H_2}}}}}

\def\fm{\hbox{$.\!\!^{\rm m}$}}

\begin{document}

   \title{Star formation in cometary globule 1: the second generation
     \thanks{Based partly on observations collected at the European
       Southern Observatory, La Silla, Chile and partly obtained from
       the ESO/ST-ECF Science Archive Facility}\fnmsep
     \thanks{Appendices A, B and C are only available in electronic
       form via http://www.edpsciences.org}}

   \author{L.\,K.  Haikala
          \inst{1,2}              
\and
          M.\ M. M\"akel\"a\inst{1,2}
\and
          P. V\"ais\"anen\inst{3,4}              
        }

 \institute {Observatory,  University of Helsinki, Finland
\and
             Department of Physics, Division of Geophysics and Astronomy,\\
             P.O. Box 64, FI-00014 University of Helsinki, Finland\\
              \email{lauri.haikala@helsinki.fi}
\and
 South African Astronomical Observatory,\\
 P.O. Box 9, Observatory 7935, Cape Town, South Africa
\and
Southern African Large Telescope Foundation,\\
 P.O.Box 9, Observatory, 7935, Cape Town, South Africa}
 \date{}

 \abstract {Cometary globule 1 (CG~1) is the archetype cometary
    globule in the Gum nebula.}  {We attempt to discover stars
    possibly embedded in the globule head and to map the distribution
    of ISM in it.}
{We analyse \ceo spectral line observations, NIR spectrosopy, narrow
  and broad band NIR imaging, and stellar photometry to determine the
  structure of the CG~1 head and the extinction of stars in its
  direction.}
{A young stellar object (YSO) associated with a bright NIR nebulosity
  and a molecular hydrogen object (MHO 1411, a probable obscured
  HH-object) were discovered in the globule.  Molecular hydrogen and
  Br$\gamma$ \ line emission is seen in the direction of the YSO.  The
  observed maximum optical extinction in the globule head is
  9\fm2. The peak N(\Htwo) \ column density and the total mass derived
  from the extinction are $9.0\times10^{21}$cm$^{-2}$ and 16.7 \Msun
  (d/300pc)$^2$. The \ceo emission in the globule head is detected in
  a 4\arcmin\ by 1\farcm5 area with a sharp maximum SW of the
  YSO. Three regions can be discerned in \ceo line velocity and
  excitation temperature.  Because of variations in the \ceo
  excitation temperature the integrated \ceo line emission does not
  follow the optical extinction.  It is argued that the changes in the
  \ceo excitation temperatures are caused by radiative heating by NX
  Pup and interaction of the YSO with the parent cloud. No indication of a strong molecular 
 outflow from the YSO is evident in the molecular line data.
The IRAS point source 07178--4429 located in the CG~1
    head resolves into two sources in the HIRES enhanced IRAS images.
    The 12 and 25\,$\mum$\ emission originates mainly in the star NX
    Puppis and the 60 and 100\,$\mum$\ emission in the YSO.}
{}


   \maketitle
%
%________________________________________________________________
\titlerunning{Star formation in Cometary globule CG~1}
\authorrunning{L.\,K. Haikala et al.}
\section{Introduction} \label{sec:introduction}

\citet{sandauleak1974} noted a 5\arcmin \ by 10\arcmin \ dark cloud west
of the emission-line star Hen 32 (\object{CoD -44 3318}, \object{NX Pup}). This
dark cloud was later included in a list of cometary globules by
\citet{HawardenBrand1976}.  Cometary globules are elongated dark
clouds with compact, dusty ``heads'' and long faintly luminous
``tails''. The first list contained 12 sources of which ten were in
the outskirts of the Gum nebula.  The sizes vary from few arcminutes
(e.g. CG~5) to tens of arcminutes (CG~12). Since their discovery
similar cometary structures have been found to be common among
interstellar clouds, the scales ranging from isolated tiny globules to
complete cometary shaped star-forming regions such as  the Corona
Australis cloud.  It has been suggested that the formation of the
classical CGs is caused by  radiation-driven implosion
\citep{reipurth1983} of an isolated, extended globule or a passage of
a supernova plane blast wave colliding with an extended globule
\citep{brandetal1983}.

\object{CG~1} \citep{HawardenBrand1976} is a prototypical cometary
globule. It has a few arcminute sized compact bright rimmed head and
nearly a half-a-degree-long tail.  The bright pre-main-sequence binary
star NX Pup lies just outside the opaque globule head. The  IRAS point
source 07178--4429 lies between NX Pup and the opaque globule head.
It has {far infrared (FIR)} colours similar to young
stellar objects.  Extensive molecular line observations of CG~1 are
presented in \defcitealias{harjuetal1990}{HSHWSP}
\citet{harjuetal1990} (hereafter \citetalias{harjuetal1990}). Even
though \citetalias{harjuetal1990} report rather strong \twco lines
(SEST antenna temperatures up to 10K), the \ceo (1--0) emission is
weak. In the \ceo maximum located in the globule head, the integrated
\ceo (1--0) main-beam brightness temperature (\Tmb) was less than 0.55
\kkms.  Two \twco velocity components are seen in the globule
head. The stronger one follows the globule optical image from head to
tail. The weaker velocity component is distributed in an elongated
structure and coincides with the bright ``nose''-like extension in
optical surface emission, which extends to below the star NX Pup. The
\citet{bourkeetal1995a} single point ammonia observation in the CG~1
head was a non-detection.  Single point observations of \object{IRAS
  07178--4429} in CO(3--2), \ceo(3--2), HCO$^+$ (4--3), and
H$^{13}$CO$^+$ (4--3) are presented in \citet{vanKempenetal2009}. The
authors conclude that the point source has no associated HCO$^+$ core
and that it very likely does not have a circumstellar shell.

     The distance to CG~1 is uncertain and the estimates range from
     300pc \citep{franco1990} to 500pc
     \citep{brandetal1983}. \citet{franco1990} analysed the
     distribution of stellar E(b--y) colour excesses in the direction
     of the Gum nebula and concluded that the distance to the centre
     of the nebula is 290$\pm$30pc. The \citet{brandetal1983} distance
     relies on NIR photometry of the variable star NX Pup, which lies
     near the globule head.  In accordance with the \citet{franco1990}
     result, \citetalias{harjuetal1990} adopted 300~pc as the distance
     to CG~1. This paper will also use the 300~pc.

It has been suggested that the Herbig AeBe star NX Pup was formed in
the globule \citep [e.g.,][]{brandetal1983, reipurth1983}. The
star is a close visual binary \citep{Bernaccaetal1993} and a further T
Tau star located 7\arcsec \ away was discovered by
\citet{brandneretal1995}.  The globule head is strongly obscured in
the optical and  in the 2MASS survey the number of stars in the
globule head is also smaller than in the surroundings. A deeper-than-2MASS
near-infrared (NIR) study is therefore needed to study the possible
stellar population still embedded in the globule and the distribution
of dust and gas in the globule head. We present  \J, \H, and \Ks \ NIR
imaging of the CG~1 head with the InfraRed Survey Facility (IRSF) 1.4m
telescope at the South African Astronomical Observatory SAAO and the
3.5m New Technology Telescope (NTT) at La Silla. Low resolution NIR
spectroscopy and narrow band imaging in the \Htwo \ 1-0 S(1) 2.12
$\mum$ line were also conducted with the NTT.  The \ceo (1--0) and (2--1)
molecular line observations of the globule head were obtained at the
Swedish ESO Submillimetre Telescope (SEST) at La Silla.  Observations,
data reduction, and calibration procedures are described in
Sect.~\ref{sec:observations_reductions} and the observational results in
Sect.~\ref{sec:results}. The new results are discussed and compared
with available data at other wavelengths in
Sect.~\ref{sec:discussion}.  The results are summarised and the
conclusions drawn in Sect.~\ref{sec:summary}.

%__________________________________________________________________

\section{Observations and data reduction}  \label{sec:observations_reductions}

\begin{figure}
\centering
\includegraphics [width=8cm] {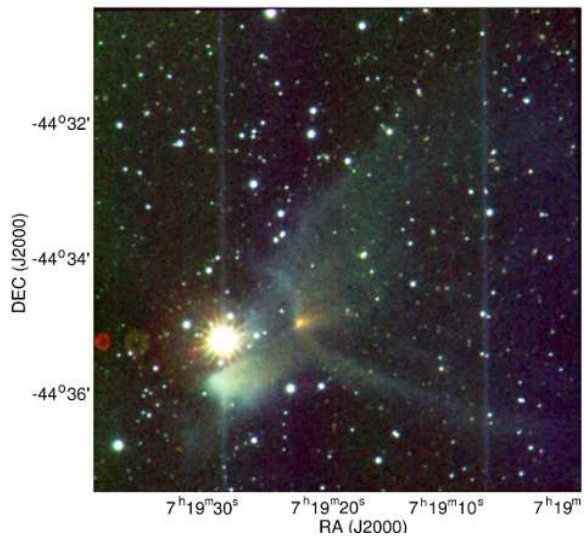}
   \caption{Colour-coded SIRIUS image of the CG~1 head. The \J, \H,
     \ and \Ks \ bands are coded in blue, green, and red,
     respectively. Square root scaling has been used to enhance
     the faint surface brightness structures. NX Pup is the brightest star 
in the image at 7$^{\mathrm h}$19$^{\mathrm m}$28\fs3, $-44^{\mathrm
    o}$35\arcmin10\farcs3} 
\label{figure:IRSF3col}
\end{figure}

\begin{figure*}
\centering
\includegraphics [width=18cm] {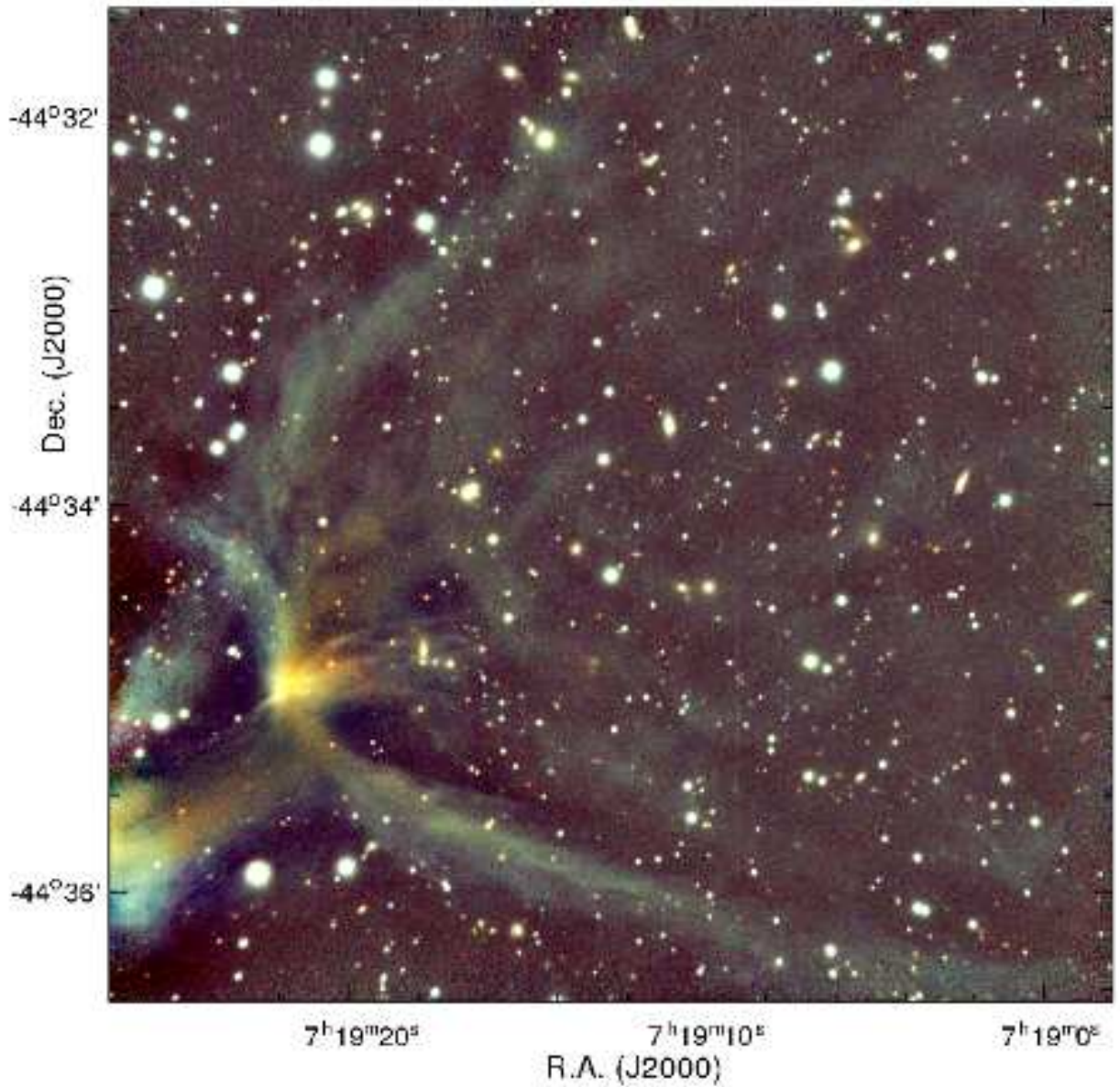}
    \caption{Colour-coded SOFI image of CG~1 head. The \Js, \H, \ and
     \Ks \ bands are coded in blue, green, and red,
     respectively. Square root scaling has been used to enhance
     the faint surface brightness structures.}
\label{figure:CG1_3col}
\end{figure*}

\subsection{NIR imaging} \label{imaging}

The head of CG~1 was imaged in \J, \H, and \Ks \ with the Simultaneous
InfraRed Imager for Unbiased Survey \citep
[SIRIUS,][]{Nagayamaetal2003} on the IRSF 1.4m telescope on
Jan. 2007. The three colours were observed simultaneously. The SIRIUS
field of view is 7\farcm7$\times$ 7\farcm7 and the pixel scale
0\farcs453.  The observations were hampered by high humidity and the
average seeing was worse than 1\arcsec.  The observations were carried
out in the on-off mode instead of the standard jitter mode to preserve
the surface brightness. After each on-integration of 10s, an off-position 
outside the globule was observed with the same integration
time. Jittering was performed after each on-off pair.
Sky flats were observed every evening and morning.

The Son of Isaac (SOFI) imaging in \Js, \H, and \Ks \ at NTT was performed
in Feb 2007. The SOFI field of view is 4\farcm9 and the pixel size
0\farcs288. The observing was done in the standard jittering mode in
observing blocks of approximately  1 hour.  The integrations consisted of 6
individual 10 second integrations and the observed total on-source
time was 1.5 hours in each filter. The average seeing was
$\sim$0\farcs7.  The bright pre-main-sequence star NX Pup, which is
located to the east of the globule head, was not within the field of view. A
reference position {at the same Galactic latitude but} outside the
globule (7$^{\mathrm h}$18$^{\mathrm m}$ -44$^{\mathrm o}$05\arcmin,
J2000) was observed in all colours. Standard stars from the \citet
{perssonetal1998} faint NIR standard list were observed frequently
during the nights.

Extended molecular hydrogen emission was searched for by imaging the CG~1 
head with the narrow SOFI filters NB H2 S1 (\Htwo \ 1-0 S(1), 2.12
$\mum$) and NB 2.090 (continuum).  The narrow-band imaging was done in the
standard jittering mode.

 The IRAF{\footnote {IRAF is distributed by the National Optical
     Astronomy Observatories, which are operated by the Association of
     Universities for Research in Astronomy, Inc., under cooperative
     agreement with the National Science Foundations}} external
 XDIMSUM package was used to reduce the SIRIUS and SOFI imaging data.
 The images were examined for cosmic-ray hits and then sky-subtracted.
 The two nearest images in time to each image were used in the
 sky subtraction. An object mask was constructed for each
 image. Applying these masks in the sky subtraction produced hole
 masks for each sky-subtracted image.  Special dome flats and
 illumination correction frames provided by the NTT team were used to
 flat field and illumination-correct the sky-subtracted SOFI
 images. Sky flats were used to flat-field the SIRIUS images.
 Rejection masks combined from a bad pixel mask and individual cosmic
 ray and hole masks were used when averaging the registered images.
The reduced SOFI \Js, \H, and \Ks \ images are shown in 
Figs. \ref{figure:online_Js}, \ref{figure:online_H}, and \ref{figure:online_Ks}.

NX Pup is very bright and produces an artifact (inter-quadrant row
cross talk) which appears as a stripe in the same column as the star
and symmetrically on the other half of the detector in the SIRIUS
images.  Reflections in the SIRIUS filters cause ring-like ghosts
around NX Pup.  We attempted  to correct for neither the inter-quadrant row
cross talk nor the reflections.

The SExtractor software v 2.5.0 \citep {bertinarnouts1996} package was
used to obtain stellar photometry of the reduced SOFI images. The
galaxies were excluded using the SExtractor CLASS keyword and by visually
inspecting the images.  After elimination of the non-stellar sources,
262 stars remained.  The magnitude zero points of the summed  data
were fixed using the standard star measurements. The instrumental
magnitudes were first converted to the \citet {perssonetal1998}
photometric system and then to the 2MASS photometric system as
described in \citet {ascensoetal2007}. The magnitude scale was checked
by comparing the SOFI photometry of stars in common with 2MASS 
that have high quality 2MASS photometry.  The limiting magnitudes
are approximately 21\fm5, 20\fm5, and 20\fm5 for \J,
\H, \ and \Ks, respectively.  The limiting magnitude however varies
over the observed area and is brighter  in the regions where the background
surface brightness is strong.

\subsection {NIR spectroscopy}

  A low resolution SOFI spectrum was acquired covering the wavelengths from 1.53
  $\mum$ \ to 2.52 $\mum$ (resolving power 980) using a  slit oriented
  along a bright surface brightness feature (see
  Figs. \ref{figure:notable} and \ref{figure:yso}). The
  SOFI standard nodding observing template was used. La Silla
  observatory IR standard F7V star Hip37123 was observed as a telluric
  standard.
 
The spectroscopic science frames were flatfielded using ON-OFF flats,
illumination and cosmic-ray corrected, and then subtracted pairwise
from each other. Xe-arcs were used to fit the wavelength solution and
correct for the two dimensional shape distortion. Frames were then
shifted, coadded, and traced.  The standard star was reduced in
exactly the same way, and its one dimensional extracted spectrum was
divided into the two dimensional target per row, and the result
multiplied by a smooth black-body model of the star, thereby removing
telluric features and also performing a relative flux calibration.

\subsection{\ceo mapping}

 The \ceo (1--0) and (2--1) molecular line observations were made in
 Sep. 2000 at SEST. The observations were conducted with the SEST 3
 and 1~mm~(IRAM) dual SiS SSB receiver using the frequency-switching
 observing mode. {A 6 MHz and 18 MHz frequency switch was used at the
   \ceo (1--0) and (2--1) line frequencies, respectively.}  The CG1
 globule head was mapped simultaneously in the C$^{18}$O~(1--0) and
 the C$^{18}$O~(2--1) transitions in a regular grid with a spacing of
 20\arcsec. Altogether 137 positions were observed using one-minute
 integration times. Selected positions were integrated longer.  Typical
 values of the effective SSB system temperatures outside the
 atmosphere ranged from 170~K to 250~K.  The SEST high-resolution 
2000-channel acousto-optical spectrometer (bandwidth 86\,MHz, channel
 width 43\,kHz) was divided into two to measure the two receivers
 simultaneously. At the observed frequencies, 109.782173 GHz and
 219.560353 GHz, the spectrometer channel width corresponds to
 $\sim$0.12~\kmps\ and $\sim$0.06~\kmps, respectively. The SEST HPBW
 and main-beam efficiency were 47\arcsec \ and 0.7,respectively,  
for the \ceo (1-0)
 frequency, and 23\arcsec \ and 0.5 for the \ceo (2--1) frequency.
 Calibration was performed using  the chopper wheel method.  Pointing was
 checked regularly towards SiO maser sources and the pointing accuracy
 is estimated to be better than 5\arcsec.

The frequency-switched spectra were first folded and a
second order baseline was then subtracted. The observed spectrum rms was
typically $\sim$0.14K and $\sim$0.2K in the \Tmb \ scale {for the
  \ceo (1--0) and (2--1) transitions, respectively}.  All the SEST line
temperatures in this paper are in the \Tmb \ units, i.e, corrected to
outside the atmosphere,  assuming that the source fills the main
beam.

\section{Results} \label{sec:results} 

\subsection{\jhks \ Imaging} \label{sec:imaging}

False colour SIRIUS and SOFI images of the CG~1 head are shown in
Figs. \ref{figure:IRSF3col} and \ref{figure:CG1_3col}, respectively.
The \J, \H, \ and \Ks \ bands are coded in blue, green, and red,
respectively. The SOFI image contains only the obscured globule head,
whereas the SIRIUS image also contains the bright pre-main-sequence
star NX Pup and a bright reflection nebulosity south of it.  The
SIRIUS observations were made in the on-off mode, which conserves the
extended surface brightness.  The SOFI observations were obtained in
the jitter mode where the surface brightness with scale larger than
the jitter box (30\arcsec \ in this case) is smeared and/or cancelled
in the data reduction. Only small-size features and gradients in the
original surface brightness structure are retained. {Point-like sources
and galaxies are unaffected.}

The edges of the globule are well defined in both the SIRIUS and the
SOFI images. In the SIRIUS image, the surface brightness is extended NW
of NX Pup and a narrow filament follows the globule southern edge.
The surface of the CG~1 head is covered with narrow, faint filaments
in the SOFI image.  Because of the jittering observing mode, the
extended surface brightness is transformed into filaments that trace
small-scale gradients in the actual surface brightness.

\begin{figure} \centering \includegraphics 
[width=8.8cm]{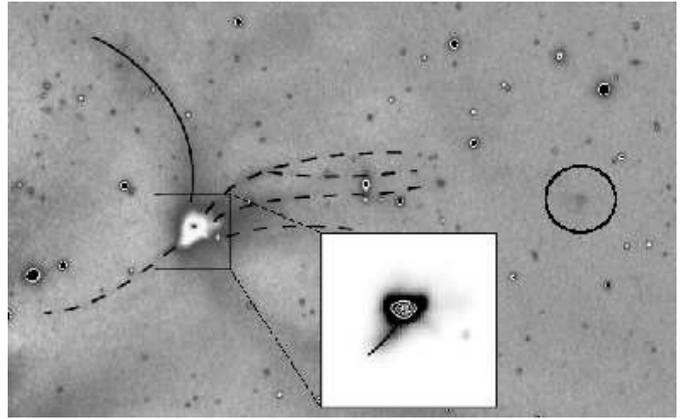}
\caption{The notable surface brightness features discussed in the text
  are marked on an extract of the SOFI \Ks \ image.  The position of
  the semi-stellar source is marked in the centre of the bright
  nebulosity (white in the figure) surrounding it. The YSO\_SE
  filament is shown with a line in the insert. The contours in the
  insert delineate the central source.  The continuous curve marks the
  ring-segment-like structure facing the NX Pup and the dashed curves
  the brightest filaments emanating from the YSO.  The faint surface
  brightness patch is inside the circle in the right.}
\label{figure:notable}
\end{figure}

{Besides the stars and numerous galaxies, a semistellar source
  surrounded by a bright nebulosity is seen in both the SIRIUS and
  SOFI images. The nebulous source has the appearance of a
  typical young stellar source \citep[see e.g.][]{zinneckeretal1999}
  and is referred to as YSO in the following.  A bright and
  elongated surface brightness feature protrudes SE of the YSO in the
  \Ks \ image.  This feature is referred as the YSO\_SE filament
  in the following.  Thin, wispy filaments extend both  west and east of
  the YSO in Fig. \ref{figure:CG1_3col}. The nebulosity associated
  with the YSO is extended and very bright. Therefore, it had to be
  masked heavily in the jitter sky frames during the data
  reduction. As a consequence, most of the structure seen in the
  reduced images around the YSO is real and  not smeared like the
  other extended structures in the SOFI images.  A faint surface
  brightness patch is seen in the SOFI images $\sim$90\arcsec \ west
  of the YSO. Both the thin filaments and the surface brightness patch
  west of the YSO are seen in all three colours.  Except for the YSO
  and the regions NW and SE of it the globule NIR surface brightness
  is predominantly blue in Fig. \ref{figure:CG1_3col}, i.e. more
  pronounced in the \Js \ band than in the two other colours.  The
  ring-segment-like structure facing the bright star NX Pup seen in
  Figs. \ref{figure:IRSF3col} and \ref{figure:CG1_3col} as well as the
  bright patch south of this star (Fig. \ref{figure:IRSF3col}) have
  counterparts in the optical images of CG~1 \citep[see e.g.][Plate~1]
  {brandetal1983}. The YSO and these features are identified in
  Fig. \ref{figure:notable}.  The ring-segment-like structure and the
  brightest thin filaments are indicated by a continuous line and
  dashed lines, respectively. The YSO\_SE filament is marked with a
  line in the insert and the nebulous patch west of the YSO with a
  circle}

\begin{figure} \centering \includegraphics 
[width=5.4cm, angle=-90]{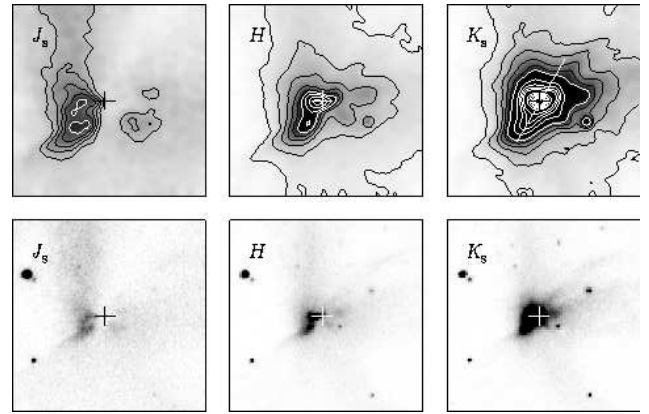}
\caption{SOFI \jhks \ images of the CG~1 YSO. The lower panels show
  the original image. The filter is indicated in the upper left corner
  of each panel. The size of the lower panels is 46\arcsec \ by
  46\arcsec. The upper panels (23\arcsec \ by 23\arcsec) show surface
  brightness contours overlayed on Gaussian smoothed images. The line
  in the upper \Ks \ panel shows the orientation of the SOFI slit on
  the nebulosity (see. Sect.\ref{sec:spectrometry}). The cross in all
  the panels indicates the position of the semistellar YSO. {The
  contour levels in SOFI counts are: \Js \  2 to 5 by 1 and
  (in white) 6; \H\ 3 to 23 by 5, (in white) 28, 36, 50, 70;
  \Ks\  3 to 23 by 5, (in white) 28 to 112 by 16, and 200 to
  600 by 80}}
\label{figure:yso}
\end{figure}

A zoomed image of the YSO in the SOFI image in three colours is shown in the
three lower panels of Fig. \ref{figure:yso}.  The filter is indicated
in the upper left corner of each panel. The size of the lower panels
is 46\arcsec \ by 46\arcsec, which approximately corresponds to the
SEST HPBW at the \ceo (1--0) line frequency, 109 GHz. The upper
23\arcsec \ by 23\arcsec \ panels (SEST HPBW at the \ceo (2--1)
frequency, 219 GHz) show surface brightness contours 
superimposed  on Gaussian smoothed images. {The contour levels and the
  grey scale were chosen to enhance the contrast in   surface
  brightness for  each filter.}  The line in the \Ks \ image shows the
orientation of the spectrometer slit.  The cross in all the panels
indicates the position of the {semi-stellar source} in the \Ks
\ image. {The \H \ and \Ks \ surface brightness contours are
  overlaid on a 23'' by 23'' \Js \ grey scale image in
  Fig. \ref{fig:YSOcontours}. The absolute surface brightness in the
  \Js \ image is very low  compared with that in the \Ks\ image. The
   \Ks \  semi-stellar source  lies behind a narrow
  obscuring lane in the \Js \ image.  In the \H \ filter, the maximum
  emission lies  east of the \Ks \ maximum and is not
  stellar like. The
  half width of the \Ks \ semi-stellar source intensity profile in the EW
  direction is nearly twice that of isolated stars elsewhere in the
  image.} This indicates that even at the \Ks\ maximum, a large part of
the observed emission is reflected light and not direct light from the
central source.

The YSO is visible in the 2MASS survey (\object {2MASS
  07192185-4434551}), which is a \Ks \ band detection (13\fm42) only. The 2MASS
extended source \object {2MASX J07192176-4434591} corresponds to the
extended YSO nebulosity.  The YSO is also visible in the \Ks \ image
presented in \citet{santos1998}, who identify it as a red object,
possibly {a  YSO in an early evolutionary stage embedded in the globule}.

\subsection{NIR spectroscopy} \label{sec:spectrometry} %

\begin{figure} \centering \includegraphics 
[width=8.8cm]{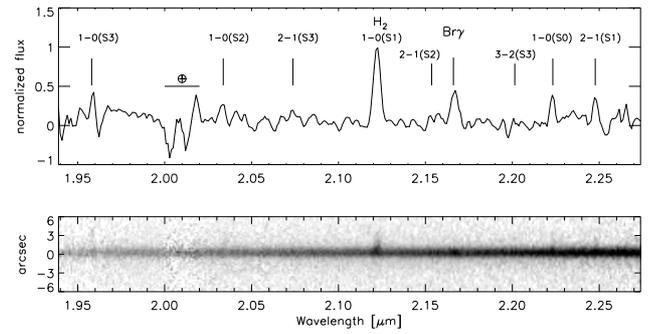}
\caption{Part of the SOFI spectrum over the YSO and its associated
  nebulosity in both two dimensional  and extracted format. Lower panel: 12\arcsec
  \ centre part of the {low-resolution long-slit spectrum through
    the YSO and the YSO SE\_filament.}  Upper panel: extracted
  continuum-subtracted spectrum convolved with a 3-pixel box-car.  The
  wavelengths of the \Htwo \ lines within the displayed spectral
  region and of the Br$\gamma$ \ line are indicated.  The spectrum has
  been normalized for an (1--0) S(1) intensity of unity. { The
  horizontal bar at 2.01 $\mum$ indicates the extent of the strong
  atmospheric absorption between the \H \ and \Ks \ bands.}}
\label{figure:sofispec}
\end{figure}

Part of the SOFI low-resolution long-slit spectrum, in both two
dimensional and extracted format, is shown in
Fig. \ref{figure:sofispec}.  The 12\arcsec \ centre part of the
spectrum through the YSO nebulosity and the YSO SE\_filament is shown
in Fig. \ref{figure:sofispec}, lower panel.  Extracted continuum-
subtracted spectrum convolved with a 3-pixel box-car is shown in the
upper panel.  The wavelengths of the \Htwo \ lines within the
displayed spectral region and of the Br$\gamma$ \ line are indicated.
The spectrum has been normalized for an (1--0) S(1) intensity of
unity. 

At least three \Htwo \ lines and the Br$\gamma$ \ line are evident in
the spectrum.  The \Htwo \ lines are detected in the centre of the
nebulosity and also at the base of the bright YSO\_SE filament. The
filament is most clearly seen in the \Htwo\ 1-0 (S1) transition.  The
Br$\gamma$ emission is strictly constrained to the spatial area of the
continuum peak.  A relatively strong continuum is seen in the
direction of the {semi-stellar YSO}.

\subsection{\Htwo \ imaging} \label{sec:h2imaging}

\begin{figure} \centering \includegraphics 
[width=7cm, angle=-90.0]{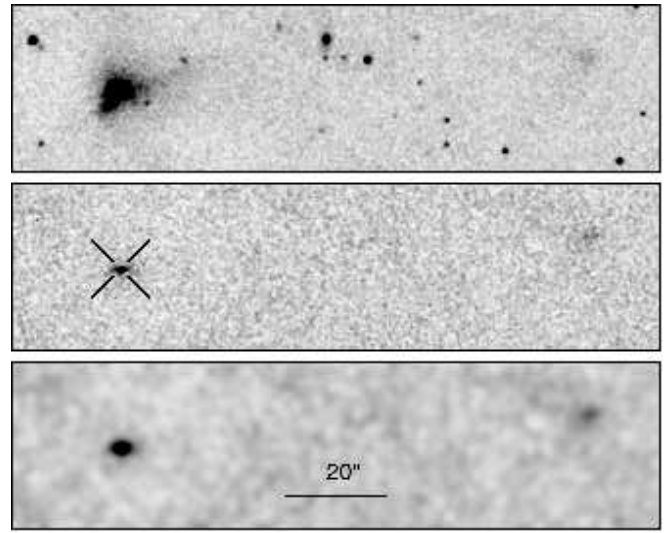}
\caption{\Htwo \ emission in the CG1 head. The upper panel shows an
  extract of the image obtained with the SOFI NB H2 S1 filter. The
  centre panel shows the difference NB H2 S1 - NB 2.090.  The cross
  shows the position of the {YSO} in the \Ks
  \ image.  The lowest panel shows the difference image smoothed with
   a three-pixel half-width Gaussian.}
\label{figure:H2}
\end{figure}

The difference between the images obtained using the narrow filter NB
H2 S1 covering the \Htwo \ 1-0 S(1), 2.12 $\mum$ \ line and the
adjacent continuum obtained with the filter NB 2.090  reveals two
objects. The stronger one lies at the centre of the YSO nebulosity and
another 90$\arcsec$ west of it. The latter coincides with the faint
surface brightness patch, which is seen in the SOFI images in all three
colours and noted in Sect. \ref{sec:imaging}. An extract of the
NB H2 S1 image and  the NB H2 S1 - NB 2.090
difference image covering the two objects is shown in the two upper panels in
Fig. \ref{figure:H2}. The cross in the middle panel shows the position
of the {{YSO}.  The stars and the
strong  surface brightness surrounding the YSO are cancelled out and
only a small size source in the direction of the YSO and a faint patch
west of it remain in the difference image.} 
The lowest panel shows the difference image smoothed by a  three-pixel 
half-width Gaussian.  

\subsection{Photometry} \label{sec:photometry} 

\begin{figure} \centering \includegraphics 
[width=8.8cm]{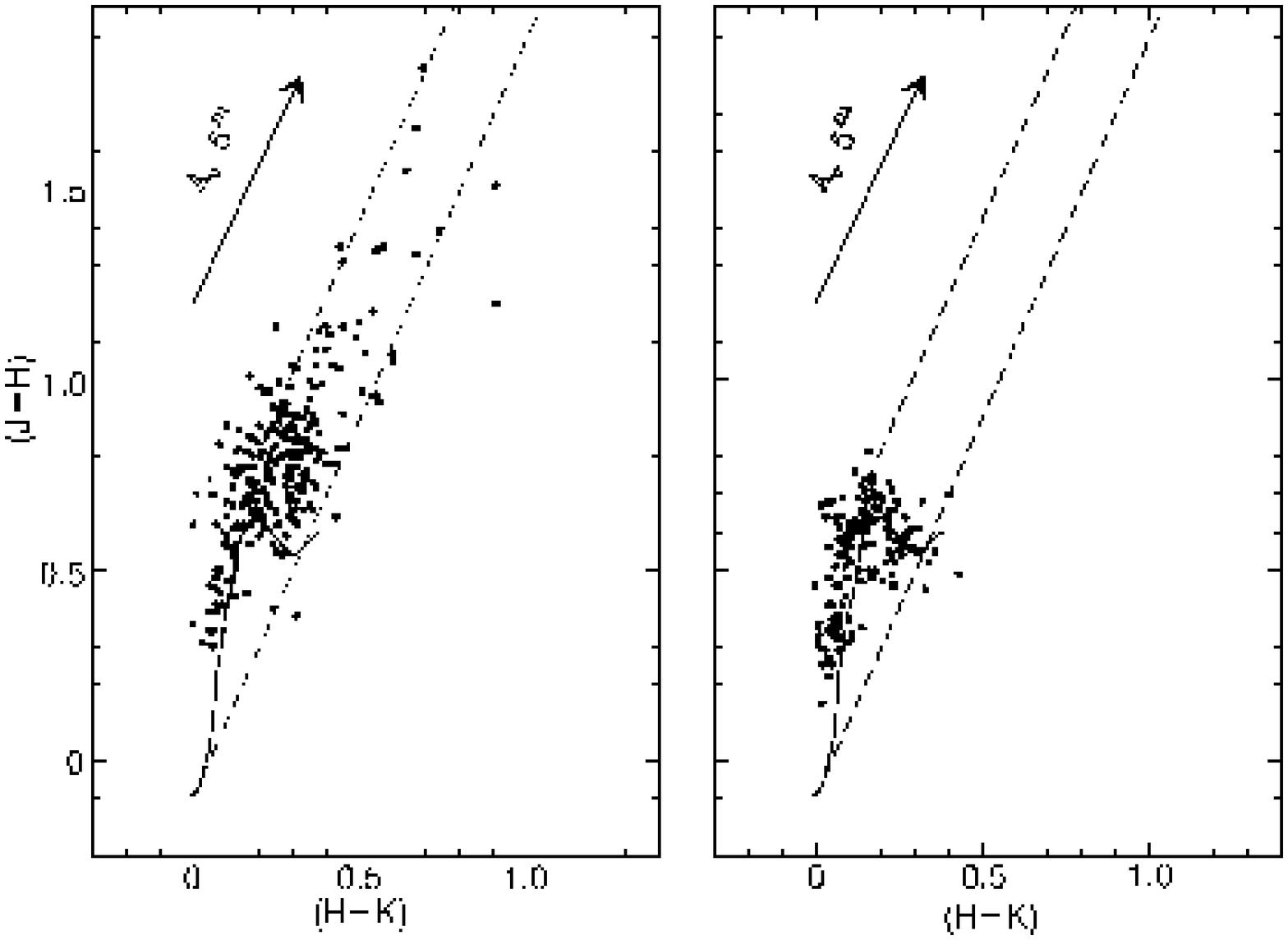}
\caption{Left: \jhks \ colour-colour diagram for stars in the CG~1 head. 
The stars with (\J-\H) and/or
  (\H-\Ks) error larger than 0\fm1 are plotted using a small size
  symbol.  The \citet{besselbrett1988} main-sequence transformed to
  the 2MASS photometric system is also plotted.  The arrow shows the
  effect of 5\umag \ of visual extinction in the diagram according to
  the \citet{besselbrett1988} reddening law. Right: As on the left, but
  for stars observed at the reference position.}
\label{figure:jhk}
\end{figure}

The \jhks \ colour-colour diagram for stars in the CG~1 head is shown
in Fig. ~\ref{figure:jhk}, left panel.  The main-sequence
\citep{besselbrett1988} transformed to the 2MASS photometric system is
also plotted.  The arrow shows the effect of 5\umag \ of visual
extinction in the diagram according to the \citet{besselbrett1988}
reddening law. A small symbol is used if the (\J-\H) and/or the
(\H-\Ks) formal error is larger than 0\fm1. The right panel shows the
colour-colour diagram for stars in the reference position. {
Normal reddened main-sequence stars should lie near or between the
reddening lines indicated with dashed lines in Fig. \ref{figure:jhk}.
Stars with circumstellar shells and disks have infrared excess and 
lie to the right of the reddening line. 
Carbon stars, long period variables, and
 extragalactic
  sources may mimic infrared excess sources \citep{hunting2008}. Even
  though the obvious extragalactic sources were filtered out from the
  original data set it is highly likely that some of these sources
  remain in the final data set. The two stars in CG~1 
that lie to the right of the
  reddening line in Fig.  ~\ref{figure:jhk} are faint
  (\Js $\sim$ 20\fm7) and were possibly  not recognised  as extended sources.
If these objects were stars associated with CG~1, they would have to be 
extremely faint (substellar). An unreddened  M5 star at a distance of 300pc
has an apparent \J \ magnitude  of 14.6.}

\subsection{\ceo mapping} \label{sec:ceo} 

\begin{figure} \centering \includegraphics 
[width=8cm]{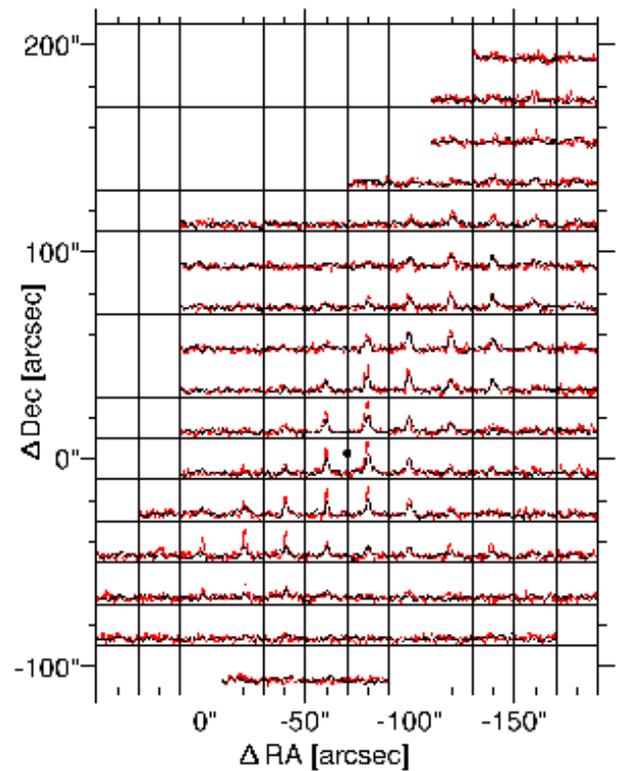}
\caption{The observed \ceo (1--0) (in black) and (2--1) (in red)
  spectra.  The spectra are plotted in the main-beam brightness scale
  from 1.0 \kmps \ to 6.0 \kmps \ in velocity and from -0.5 K to 2.8 K
  in temperature. The positional offsets are in arcseconds from
  7$^{\mathrm h}$19$^{\mathrm m}$28\fs3 -44$^{\mathrm
    o}$35\arcmin6\farcs8 (J2000), which is 4\farcs5 north of NX
  Pup. The filled circle indicates the position of the
  YSO in the plot.}
\label{figure:ceo2-1-0}
\end{figure}

The observed \ceo (1--0) (in black) and (2--1) (in red) spectra are
shown in Fig. \ref{figure:ceo2-1-0}.  The spectra are plotted in the
main-beam brightness scale from 1.0 \kmps \ to 6.0 \kmps \ in velocity
and from -0.5 K to 2.8 K in temperature. The positional offsets are
arcseconds from 7$^{\mathrm h}$19$^{\mathrm m}$28\fs3 -44$^{\mathrm
  o}$35\arcmin6\farcs8 (J2000), which is 4\farcs5 North of NX Pup.
The map centre position is the same as in \citetalias{harjuetal1990}
molecular line maps. Contour maps of the \ceo (1--0) and (2--1)
emission are shown in Figs. \ref{figure:online_1-0} and
\ref{figure:online_2-1}.  The distribution of the \ceo (1--0) emission
in Fig. \ref{figure:ceo2-1-0} and Fig. \ref{figure:online_1-0} is
consistent with that shown in \citetalias{harjuetal1990}, Fig. 1. The
differences can be attributed to the coarse sampling used by
\citetalias{harjuetal1990} (40\arcsec\ compared to the 20\arcsec
\ spacing in this paper). In addition the low value of the
\citetalias{harjuetal1990} maximum integrated \ceo (1--0) line
integral, which is less than 50\% of the present data, may be
explained in a similar way.

\begin{figure} \centering \includegraphics 
[width=8cm]{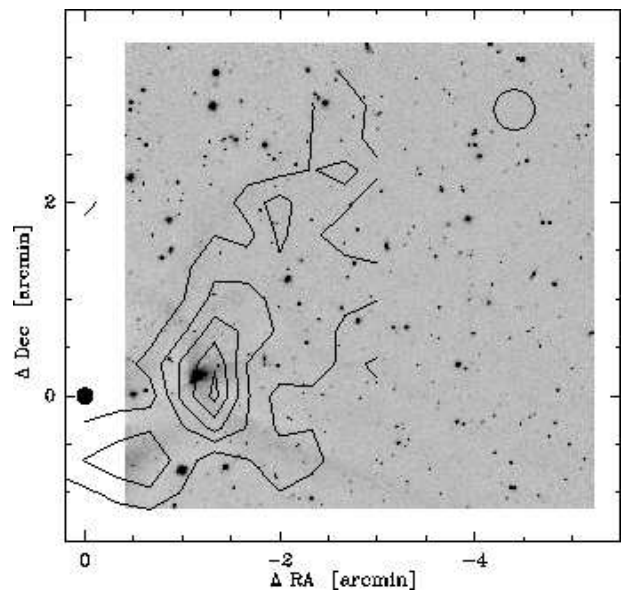}
\caption{Contour map of the \ceo (2--1) integrated emission (main-beam
  brightness scale) overlaid on the \Ks \ SOFI image. Contour levels
  are from 0.4~\kkms \ to 2.4~\kkms\ in steps of 0.4~\kkms.  {SEST
    HPBW at the \ceo (2--1) frequency is indicated in the upper right.
    The positional offsets are in arcminutes from NX Pup indicated
    with a filled circle.}}
\label{figure:ceo_int}
\end{figure}

A contour map of the integrated \ceo (2--1) emission in the main-beam
brightness scale superimposed  on the SOFI \Ks \  image is shown in
Fig. \ref{figure:ceo_int}. The contours are from 0.4~\kkms \ to
2.4~\kkms\ in steps of 0.4~\kkms. The distribution of the emission is
elongated with a sharp maximum SW of the  YSO.
The general \ceo distribution correlates well with
the optical extinction evident in the SOFI \Js\ image (Fig.  \ref{figure:online_Js}).  
 
\begin{figure} \centering \includegraphics 
[width=8.8cm]{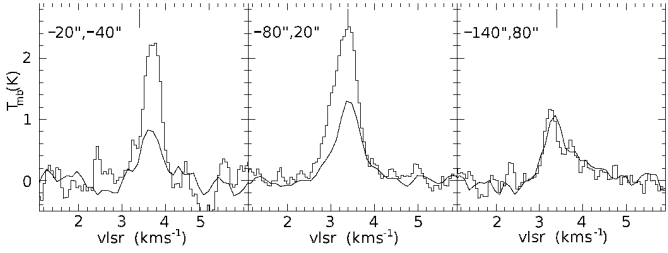}
\caption{\ceo (2--1) (histogram) and (1--0) (line) spectra
in three positions. The offset from the map centre is indicated in the
upper left corner of each plot. The velocity marker at the top
of each panel is at velocity 3.4 \kmps.}
\label{figure:specceo}
\end{figure}

The  C$^{18}$O (1--0) and (2--1) spectra at three positions are shown in
  Fig.  \ref{figure:specceo}. The offset from the map centre position
  is shown in arcseconds at the upper left corner of each panel. The
  velocity marker at the top of each panel is at velocity 3.4
  \kmps. The offset positions (-20\arcsec,-40\arcsec),
  (-80\arcsec,20\arcsec) and (-140\arcsec,80\arcsec) are  referred
  to as SE, centre and NW positions, respectively.
 In Fig. \ref{figure:specceo}, the \ceo (2--1) line
  intensity is significantly stronger than the \ceo (1--0)
  intensity at the  SE and centre positions.
  However, the {centre of line velocities }are different,
3.69$\pm$0.03~\kmps\  at SE and  3.33$\pm$0.03~\kmps \ at  the
centre position.  Also the line widths differ being 0.46~\kmps \ and 
0.70~\kmps\  at  the two positions, respectively.  
  The velocity difference was seen by
  \citetalias{harjuetal1990} in \twco and here it is confirmed in the
  \ceo emission. At the NW position,  the intensities in the two
transitions 
  are equal and the line velocity is similar to that at the
centre position. The lines are asymmetric (blue wing at centre position,
and red at NW position).

\section{Discussion} \label{sec:discussion} 
 
\subsection{\ceo mapping} \label{sec:ceo_disc}

In the following discussion, the \ceo map is divided into three
regions: 1) \ceoSE \ (the 3.7 \kmps\  velocity component is dominant);
2) \ceomax \ (the integrated \ceo emission maximum); and 3) \ceoNW
\ (region in the NW where the \ceo (1-0) and (2--1) \Tmb \ intensities are
similar).

{In the LTE approximation, the \ceo line intensity depends on both 
  the \ceo column density and the \ceo excitation temperature, whereas the
  \ceo (1--0) to \ceo (2--1) intensity  ratio depends on the excitation
  temperature.}  If the \ceo excitation temperature were constant in
the mapped area, the integrated line area of the optically thin \ceo
emission would trace the cloud gas column-density linearly. However,
the \ceo peak \Tmb(2--1)/ \Tmb(1--0) ratio varies over the map. In
Figs. \ref{figure:ceo2-1-0} and \ref{figure:specceo}, the \ceo (2--1)
emission is significantly stronger than the \ceo (1--0) emission in
the direction of the \ceomax\ and in \ceoSE.  In \ceoNW, the intensity
of the two transitions is the same.  {The varying line ratio
  indicates that the excitation temperature is higher in both \ceomax \ and
  \ceoSE \ than in \ceoNW. }

The interaction of the newly born star with the surrounding cloud (see
the discussion below) would be a likely reason for a warmer spot in
the cloud (\ceomax). The \ceo line width is also the broadest in \ceomax,
which is indicative of extra turbulence at this position.
The  \ceo line velocity in \ceoSE \ differs  from that in \ceomax,
implying  that these are separate structures in the cloud.  The high
NIR surface brightness around \ceoSE \ in Figs. \ref{figure:IRSF3col}
and \ref{figure:CG1_3col} indicates that the radiation from the star
NX Pup is the reason for the elevated temperature. \ceoNW \ lies NW
of the YSO and  the \ceomax \ and is most probably well shielded against
the radiation from NX Pup. This region of the cloud would correspond
to a quiescent, low temperature part of the cloud.

 {The \ceo (2--1) emission is distributed in a
   north-south oriented ridge, which is inconsistent with a
   point-like source.
   It can be seen in Figs. \ref{figure:ceo2-1-0} and \ref {figure:ceo_int} 
   that the \ceo
   (2--1) maximum does not fall on the measuring grid but lies
   slightly NE of the measured maximum line integral and west of the
   location of the YSO. If the \ceo maximum were point-like, it would
   be justifiable to argue that the positional offset between the
   maximum and the YSO would be caused by a larger than expected
   telescope pointing error during the observations. However, the
   maximum is elongated and there is no obvious reason why the
   north-south oriented \ceo ridge should be centred on the YSO
   position.  The \ceo mapping grid spacing is only marginally finer
   than the SEST beam HPBW at the \ceo (2--1) frequency (grid
   20\arcsec, HPBW 23\arcsec) and therefore a fully sampled map is
   needed to pinpoint the exact location of the maximum.  }

{The \hcopl (4--3) line observed by \citet {vanKempenetal2009} in
  the direction of the YSO is weak (\Tmb \ 0.6K) and \htcopl (4--3) was
  a upper limit of 0.08K. This rules out a dense core in this
  direction \citep {vanKempenetal2009}.  Even though the position
  where \citet {vanKempenetal2009} observed the \ceo (3--2) and the
  \hcopl (4--3) lines, the YSO, lies between the mapping positions in
  this paper, their measurement can be used to evaluate the nature of
  the \ceomax.  The APEX beam size at 319 GHz is similar to the SEST
  beam size at 220 GHz, which makes comparing the line intensities
  observed with APEX and SEST realistic.  The APEX \ceo (3--2) \Tmb
  \ line intensity, 2.3K, is similar to the SEST \ceo (2--1) line
  intensities observed around the YSO. Because the \ceo (3--2) line
  temperature is low, the \ceomax \ position cannot be a \ceo hotspot
  similar to that discovered in CG~12 (NGC 5367)
  \citep{haikalaetal2006}. Even though the emission from high density
  tracers is weak in the CG~12 \ceo hotspot, the \ceo (3--2) \Tmb \ of
  10K is high compared to the \Tmb \ temperatures in the (2--1)
  and (1--0) transitions (5~K and 2.2~K, respectively).  Observations
  of the high density tracers \citep {vanKempenetal2009} and \ceo thus indicate
  that \ceomax \ is a moderately dense elongated structure west
  of the YSO position.}

\subsection{Visual extinction } \label{sec:extinction}

\begin{figure} \centering \includegraphics 
[width=8cm]{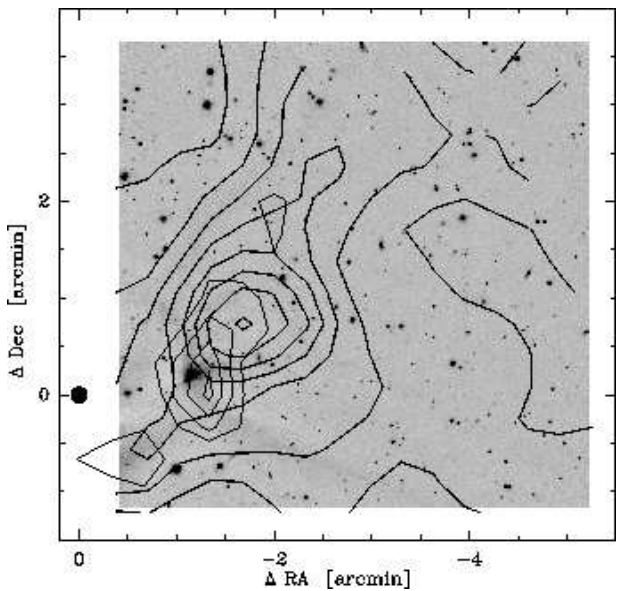}
\caption{Contour map of the optical extinction {(thick lines)} 
obtained with the NICER method
 overlaid on the \Ks \ SOFI image. Contour
  levels are from 2 magnitude to 9 magnitudes in steps of 1 magnitude.
{  Contours of the \ceo (2--1) integrated emission
(0.8 \kkms \ to 2.4 \kkms \ in steps of 0.4 \kkms) are shown with thin lines.}
  The positional offsets are as in Fig. \ref {figure:ceo_int}}
\label{figure:av}
\end{figure}

The NICER method presented in \citet{nicer} and the SOFI NIR
photometry can be used to estimate the visual extinction within the
imaged area in CG~1. However, the position of CG~1 14$^{\mathrm o}$
below the Galactic plane is unfavourable for applying the NICER
method because the expected number of bright background stars, especially
of giant stars, is low. Therefore the spatial resolution of the
resulting extinction map is only 20\arcsec.  The \J-\H/\H-\Ks
\ colour-colour diagram for the observed off-field is shown in
Fig. \ref{figure:jhk}, right panel.  {The diagram shows that the
  interstellar extinction is small in the off-field direction.  The
  field lies at the same Galactic latitude as the on-field and
  therefore the stellar population should be statistically similar to
  that of the on-field. Thus the off-field is well suited to being  used
  in the NICER method.} The contour map {(thick lines) of the
extinction derived for the CG~1 head superimposed  on the SOFI \Ks \ image
is shown in Fig. \ref{figure:av}. The contours are from 2\fm0 to 9\fm0
in steps of 1\fm0. he contours of the \ceo (2--1) integrated emission
(0.8 \kkms \ to 2.4 \kkms \ in steps of 0.4 \kkms) are shown with thin lines.}
The minimum visual
extinction, 1\fm3, in the map is in the NE corner of the image.  This
non-zero minimum extinction agrees with Fig. \ref{figure:jhk},
where practically all late main sequence stars are above the 
main-sequence also plotted in the figure. The maximum extinction, 9\fm2, is
located NW from the YSO.  {However, a word of caution is required because the
   NICER method assumes that the extinction is constant in each
  NICER cell (in this case 20\arcsec \ by 20\arcsec). If a cell
  contains a compact (with respect to the cell size) high extinction
  structure, the measured extinction tends to be biased towards lower
  extinction \citep {lombardi2005}. Thus, localised small-size
  extinction maxima may go unnoticed.}

The position of the maximum extinction does not coincide with the
maximum \ceo integrated emission (Fig. \ref{figure:ceo_int}). This
shows, as discussed in Sect. \ref{sec:ceo_disc}, that {because of
  the varying \ceo excitation temperature the distribution of the
  observed} \ceo integrated emission does not linearly trace the
gas/dust column density distribution. {In addition, the optical
  extinction also traces the low density cloud envelope where \ceo
  molecule is not excited.}

\subsection{Column densities and masses} \label{sec:masses}

{A detailed three-dimensional non-LTE model that accounts for both
  the varying density and excitation conditions would be necessary to
  derive the cloud physical properties.  The construction of this model
  is not possible because only \ceo (1--0) and (2--1) mapping data can
  be used and  only a few pointed observations are available for other
  molecules.  The LTE-approximation approach is used instead of a
  sophisticated model to obtain a zeroth-order estimate of the \Htwo
  \ column densities and masses. An average \ceo excitation
  temperature for \ceoSE, \ceomax, and \ceoNW \ is estimated assuming
  optically thin \ceo emission and using the observed relative \ceo
  (2--1) and (1--0) line intensities.  This assumes that the observed
  emission in both transitions originates in the same volume of gas at
  a constant excitation temperature.  The observed \ceo line
  \Tmb(2--1)/ \Tmb(1--0) ratio is 2 in \ceoSE \ and \ceomax, which is
  compatible with an excitation temperature near 15~K. In \ceoNW, the
  ratio is one indicating an excitation temperature of the order of
  10~K.  The low \ceo (2--1) \Tmb \ temperature in \ceoNW \  may also
  be due to subthermal excitation. However, according to model
  calculations presented in \citet{warinetal1996}, the deviation of the
  populations in the lowest \ceo energy levels from their LTE values
  is small in typical conditions prevailing in dark clouds.
  Considering the high \Htwo \ column density derived for \ceoNW
  \ from optical extinction (see below), subthermal excitation is 
  unlikely.}

The N(\Htwo) column density can be estimated using both the \ceo data
and the optical extinction. The relation between the molecular
hydrogen and \ceo column density is N(H$_2$) =
[N(\ceo)/1.7$\times 10^{14}+1.3]~10^{21}$ \citep{frerkingetal1982}. Assuming
LTE and \ceo excitation temperature of 15~K in \ceoSE \ and \ceomax
\ and 10~K in \ceoNW, the peak N(\Htwo) column densities in these
regions are $5.0\times 0 10^{21}cm^{-2}$, $7.8 \times 10^{21}cm^{-2}$, and $4.9
\times 10^{21}cm^{-2}$, respectively.  The
relation between the optical extinction and hydrogen column density is
N(H) = $2 \times 10^{21}$cm$^{-2}$mag$^{-1}$ \citep {bohlinetal1978}. The
peak N(\Htwo) column density estimated from the extinction data is
$9.0 \times 10^{21}cm^{-2}$. The maximum is in the direction of \ceoNW \ and
is nearly twice that estimated from the \ceo data.

{The LTE column densities can be compared with those calculated
  with a non-LTE radiative transfer code such as  RADEX \citep {radex2007},
  which is available on-line. The \ceo (1--0) and (2--1) line
  intensities are calculated by providing  the kinetic temperature, \Htwo
  \ number density, the \ceo column density, and the \ceo line width as
  input parameters.  The line width is known in CG~1 and a \Htwo
  \ number density of $~10^{4}$cm$^{-3}$ is a reasonable guess. The
  observed \ceo (2--1) to \ceo (1--0) line intensities in \ceoNW \ can
  be reproduced with RADEX assuming a \ceo 10K excitation temperature
  and the LTE column density calculated above. However, in \ceoSE \ and
  \ceomax \ the \ceo excitation temperature has to be increased from 15K
  to 30K to make the line intensities agree. This is because of  
  subthermal excitation of the \ceo (2--1) transition. Increasing the
  N(\Htwo) column density makes the \ceo (1--0) line stronger  in respect
  to the \ceo (2--1) line which does not agree with the observations. 
The comparison of LTE and non-LTE results shows that the
two methods produce similar  N(\Htwo) column densities if the input
kinetic temperature is adjusted. The conversion factor from the
line integral to column density in the LTE method is temperature 
dependent but the variation is only 20\% in the 10K to 30K range.
}

 The cloud mass can be calculated from the \ceo data by summing up the
 calculated N(H$_2$) column densities point by point and using the
 mean molecular weight per H$_2$ molecule 2.8.  A major uncertainty in
 the mass calculation is the distance to CG~1, as the mass scales as the 
 square of the distance. If the distance d to CG~1 is 450pc instead of  the 
assumed 300~pc, the calculated mass more than doubles. 
Assuming \ceo excitation temperatures as
 above, the masses of \ceoSE, \ceomax, and \ceoNW \ are
 0.75\Msun$~(d/300pc)^2$, 1.35\Msun$~(d/300pc)^2$,  and
 1.6\Msun$~(d/300pc)^2$, respectively.  These numbers should be
 considered very rough estimates because the regions  overlap and
 the division of data between them is subjective.  The masses are
 lower limits because they refer to gas traced by the \ceo
 emission. The true  masses are higher as the clumps have low density
 envelopes that are not detected in \ceop.  The most massive of the
 three regions is \ceoNW.  The \ceo line intensity in its direction is
 lower than in the other two  regions but this is compensated by its
 larger size.

 Summing up the extinctions point by point, using the \citet
 {bohlinetal1978} relation between the extinction and hydrogen column
 density and mean molecular weight per H$_2$ molecule 2.8, the total
 mass from the extinction data in the imaged area is
 16.7\Msun$~(d/300pc)^2$. The A$_v$= 4~magnitudes contour covers
 approximately the area observed in \ceo. The mass within this contour
 is 9.2\Msun$~(d/300pc)^2$, which is more than twice the
 summed up \ceo mass. 
The discrepancy between the masses calculated from the extinction data and
the \ceo data is expected because  the
 optical extinction also traces  the low density cloud envelope, which
 is not detected in \ceop.

\subsection{Spectroscopy and narrow-band imaging } \label{sec:spec-narrowband}

The SOFI spectrum was taken to investigate the nature of the
nebulosity, i.e., ascertain   whether it is only reflected light or whether line
emission is also present. The spectrum in Fig. \ref{figure:sofispec}
shows an underlying continuum, at least three \Htwo \  lines, and the
Br$\gamma$\ line.  The \Htwo \ line emission does not only originate in 
the location of the {YSO} but also from the base
of the YSO\_SE filament.  The spectral resolution and the spectrum
low signal-to-noise ratio do not allow us to extract detailed physical
information, though by comparing with the models of e.g.  \citet{smith1995}
we note that the line strengths of the \Htwo\ lines appear more
consistent with C-type shocks than J-type.  As expected, the
Br$\gamma$ \ line emission is seen only in the direction of the YSO as
this line is expected to originate very near the YSO \citep [see e.g.]
[] {malbet2007}. The underlying continuum emission is direct
light from the YSO and light reflected from the surroundings. The
spectrum is not sensitive enough to show the continuum in the
direction of the SE filament.

The \Htwo \ emission region can also be seen in the narrow-band
\Htwo/continuum difference image in Fig. \ref{figure:H2}. It is
centered on the \Ks \ {YSO} and is resolved in the E-W direction
but not in the N-S direction. The extent in the E-W direction is
5\arcsec.  The NIR spectrum shows that \Htwo \ emission comes also
from the base of the YSO\_SE filament but the narrow-band images are
not deep enough and emission in Fig.  \ref{figure:H2} is confined only
to the YSO.

For the offset 90\arcsec \ west from the YSO, a faint patch of emission is
detected.  This patch coincides with those seen in the \Js\ and \H
\ bands (Sect. \ref{sec:imaging}).  The difference image in Fig.
\ref{figure:H2} shows that in the \Ks \ band this emission is mostly
due to \Htwo \ line emission. Though this object is likely to be a
HH-object it cannot be confirmed in the optical. Therefore it will be
designated as a molecular hydrogen emission-line object 
\object{MHO 1411}, \citep{MHO2010}.

\subsection{IRAS 07178--4429} \label{sec:compare}  

CG~1 was mapped  by the IRAS satellite in wavelengths 12 to 100
$\mum$.  Besides  extended surface emission, IRAS detected a point source,
IRAS 07178--4429, at the edge of the globule head. 
The non-colour-corrected fluxes of the point source are 6.68, 7.60,
13.12, and 33.59 Jy at 12, 25, 60, and 100 $\mum$, respectively.  
The positional uncertainty ellipse major and  minor 
axis are 10\arcsec, 3\arcsec and the  position angle (East through North) 
178\degr. The 60 and 100  $\mum$ \  fluxes point to an embedded source.
However, considering the  100$\mum$ \  flux of 33.59~Jy, the 12  $\mum$ \  
flux of 6.68~Jy is too strong for such a source.

\begin{figure} \centering \includegraphics 
[width=8cm, angle=-90.0]{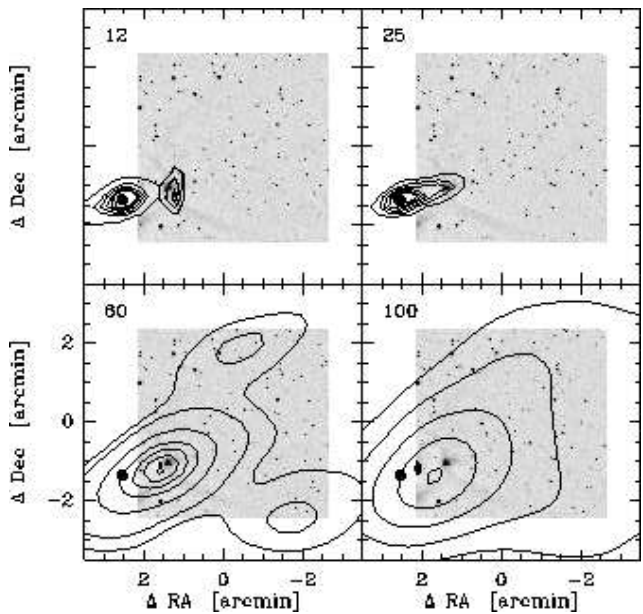}
\caption{Contour maps of the HIRES processed IRAS maps superimposed on
  the SOFI \Ks \ band image. The IRAS wavelength is shown in microns
  in the upper right corner of each panel. The position of the star NX
  Pup is marked with a filled circle.  The IRAS 07178--4429 point
  source positional uncertainty is shown as an {filled ellipse west of
    NX Pup} in the 100 $\mum$ \ panel. The \ceo (2-1) integrated
  emission contours (indicated by thick lines, 1.2 \kkms \ to 2.0
  \kkms \ in steps of 0.4 \kkms) are superimposed on the 12 $\mum$
  \ panel. The offset is in arcminutes from the centre of the SOFI
  image.{ The contour levels in MJy sr$^{-1}$ \ are: 12 $\mum$ 5 to 30
    by 5; 25 $\mum$ 15 to 115 by 20; 60 $\mum$ 3, 5, 15 to 91 by 19;
    and 100 $\mum$ 7, 16, 31 to 71 by 20.}}

 \label{figure:KsHIRES}
\end{figure}

The spatial resolution of original IRAS images can be
enhanced using HIRES processing, which uses the maximum correlation
technique \citep{aumanetal1990}.  Figure \ref{figure:KsHIRES} displays
contour maps of the CG1 HIRES processed (20 iterations)  IRAS maps 
superimposed on the SOFI \Ks \ band image. The IRAS wavelength
in microns is shown in the upper left corner of each panel and the
position of the star NX Pup is marked with a filled circle.  The IRAS
07178--4429 point-source positional uncertainty ellipse is shown in the
100 $\mum$ \  panel. It lies west of NX Pup at the edge of the \Ks \ image.
The \ceo (2-1) integrated emission contours (1.2
\kkms \ to 2.0 \kkms \  in steps of 0.4 \kkms) are superimposed on the 12
$\mum$ panel. 
The 12 $\mum$ \  IRAS contours in Fig. \ref{figure:KsHIRES} are centred
on the star NX Pup. At 25 $\mum$, the maximum contours are still
centred on NX Pup but at a lower level the contours encircle the YSO.
 At 60 and 100 $\mum$, the maximum of the 
emission has shifted significantly to the west of NX Pup and lies
near the  YSO and the maximum of the \ceo (2--1)
integrated emission. The nominal point source position lies between
the 12-25  $\mum$ \  and the 60-100 $\mum$ \  maxima.

The likely explanation of the disagreement between the point source fluxes 
(hot or
cold source) and the systematic shift of the {position of the}
IRAS emission maximum from 12 $\mum$ \ to 100 $\mum$ \ is that there
are two sources, one warm (NX Pup) and one cold (the YSO). 
  {Spitzer 24 and 70 $\mum$ \ MIPS observations now available online
  confirm the HIRES analysis.  Spitzer detected two sources, one
  centered on NX Pup and another at the YSO. Their relative flux ratios
  (YSO/NX Pup) are $\sim$0.64 (24$\mum$) and $\sim$100 (70$\mum$).}  
A similar case of two
  infrared sources observed as one IRAS point source is observed in 
    \object {CG~12} (\object{NGC 5367}). The Herbig AeBe binary star
    \object{h4636} and an adjacent cold source are merged into a
    single source, \object{IRAS 13547-3944}
    \citep{haikalareipurth2010}.

{The IRAS 12, 25, 60, and 100$\mum$ point source catalog fluxes in
  janskys can be converted into in band fluxes by multiplying them with
  the synthesized IRAS bandpasses of $2.06 \times 10^{13}$~Hz, 
$7.56 \times  10^{12}$~Hz, $4.55 \times
  10^{12}$~Hz, and $1.74 \times 10^{12}$~Hz, respectively \citep
  {emerson1988} and the source luminosity can be calculated.  {
Though the Spitzer 70$\mum$\ and the IRAS 60$\mum$\  bandpasses are not the
same, it can be assumed that the contribution of NX Pup to the IRAS 60 $\mum$ \
flux is not significant. Using the IRAS 07178--4429  60 and 100$\mum$ fluxes 
the FIR
  luminosity of the YSO  is
  3.1~\Lsun$\times (d/300pc)^2$. }}

There are two secondary maxima in addition to the IRAS point source at
60$\mum$ \ in Fig. \ref{figure:KsHIRES}. These lie at the globule
bright edge seen in Fig. \ref{figure:CG1_3col}. In the three-component
dust model \citep{pugetleger1989}, interstellar dust consists of large
aromatic molecules (PAHs), very small grains (VSGs), and the classical
large grains. The VSGs are transiently heated and emit non-thermal
emission in the mid- to FIR. In dark clouds (temperatures $<$20K), the
large grains emit in thermal equilibrium at wavelengths
$\geq$80$\mum$.  The two secondary 60$\mum$ maxima in
Fig. \ref{figure:KsHIRES} are therefore most likely to be produced by
 VSGs at the
surface of the globule and heated by the star NX Pup or the UV
radiation from the central part of the Gum nebula. {Consistent
  with the HIRES 60$\mum$ image the globule bright edges can also be
  seen in the Spitzer 70 $\mum$\ image. In addition, the  bright
reflection nebulosity at the SE tip of CG~1 seen in Fig.
\ref{figure:IRSF3col}
  appears as a bright, extended source in the Spitzer 70
  $\mum$\ image. This extended emission is likely to be due to VSGs
  heated by NX Pup. This would also agree with the argument presented
  in Sect. \ref{sec:ceo_disc} that the radiation from NX Pup explains
  the elevated \ceo excitation temperature in \ceoSE. The
  Spitzer 70 $\mum$ \ integrated flux from \ceoSE \ is similar to that
  of NX Pup and does not invalidate the YSO IRAS FIR luminosity
  calculation presented above.}

{
\subsection{NX Pup--cloud interaction} \label{sec:NXPup_interact}

The HAEBE star NX Pup lies north of the bright ``nose'' nebulosity and
east of the obscured  CG~1 globule head. Being close to CG~1, the star  NX Pup 
dominates the local radiation field over the general interstellar
radiation field at least around the globule head.
Two surface brightness features in CG~1
indicate that the position of the star near the globule is not a
projection effect but that the star is physically located near the
globule.  It is likely that the illuminating source of the very bright
reflection nebulosity at the SE tip of CG~1 seen in Fig.
\ref{figure:IRSF3col} {and the corresponding Spitzer 70 $\mum$ \ surface
brightness} is NX Pup. The star may also be the  reason
for the elevated \ceo excitation temperature of \ceoSE.  The second
surface brightness feature is the arc reaching NE from the YSO. The arc is
symmetric with respect to NX Pup and  is also seen in the optical
images, not only in NIR. The arc could indicate the edge of a bubble
vacated from gas and dust by NX Pup. 
}

\subsection{YSO properties} \label{sec:YSO}

{In the SOFI imaging, the YSO is visible only at \Ks. In \H \ and
  \Js, it lies behind an obscuring dust lane (Figs. \ref{figure:yso}
  and \ref{fig:YSOcontours}). Therefore its (\J-\H) and (\H-\K)
  colour indices cannot be calculated and its position in the \jhks
  \ colour-colour diagram is not known. The YSO 2Mass \Ks \ magnitude is
  13\fm42, which is consistent with the SOFI photometry. The SOFI
  imaging \H \ limiting magnitude is 20\fm5 but because the YSO lies
  at the centre of a very bright nebulosity the limiting magnitude is
  brighter at its position. A conservative estimate of the minimum (\H-\K)
  colour index would be $\gtrsim$6 magnitudes. The average intrinsic
  (\H-\K) colour index for the SOFI off-field in Fig. \ref{figure:jhk}
  is 0\fm15.  The YSO visual extinction calculated with the NICE
  method \citep{ladalada1994} would be $\sim$93 magnitudes. This is
  however an upper limit as the object most probably has an infrared
  excess (the extent of the YSO source profile in the EW direction is
  nearly twice that of isolated stars elsewhere in the image
  (Sect. \ref{sec:imaging}). Despite the IR excess, the visual
  extinction towards the YSO is likely to be more than 50
  magnitudes. This extinction is significantly higher than the maximum
  extinction in the CG~1 head (9\fm2) and the estimated extinction in
  the direction of the YSO ($\sim$6 magnitudes) as estimated with the
  NICER method in Sect.  \ref{sec:extinction}. If the extinction takes
  place in a compact circumstellar disk or a circumstellar shell it
  would not be detected with the NICER method.

 The \citet {vanKempenetal2009} pointed molecular line observations of
 the IRAS 07178--4429 point source were aimed at detecting the
 possible associated protostellar envelope/core and molecular outflow.
 The position observed by \citet {vanKempenetal2009} was however the
 YSO and not the nominal IRAS point source position. This pointing was
 possibly selected using the CG~1 \Ks \ image in
 \citet{santos1998}. \citet {vanKempenetal2009} found evidence of neither protostellar envelope nor a molecular outflow and
 concluded that it is most likely not an embedded YSO (class I
 source).  This indicates that the observed high extinction towards
 the YSO takes place in a circumstellar disk and not in a
 circumstellar shell. This makes the source a likely class II source.

}

\subsection{YSO--cloud interaction: where is the outflow?} \label{sec:interaction}

Besides mass infall, the early stage of low-mass star formation is also
associated with energetic mass outflows. Manifestations of outflows
are observed, e.g., as jets, molecular outflows and Herbig-Haro
objects. The outflows inject momentum and energy into the parent cloud
and can have a highly disruptive effect on the cloud core they are
forming in. 

{Molecular hydrogen emission detected in the low-mass
  star-formation regions is usually associated with shocks due to hot and
  dense gas flowing out  from the newly born stars. The \Htwo \ 
  NIR  line emission can also be due to UV absorption and
  fluorescence \citep{blackdalgarno1976}.  Fluorescent \Htwo \ line emission
  has been observed in elephant trunks in HII regions \citep[e.g.~][]
  {allenetal1999}. The fluorescent \Htwo \ emission appears  as a
  bright, thin layer on the surface of the trunks on the side facing
  the ionizing source.  However, the ionizing source is missing in
  CG~1. The FUV spectrum of NX Pup is most consistent with a spectral type F2III
\citep {blondel2006}. Thus the UV flux from NX Pup cannot  ionize 
the cloud surface. 
It is also not plausible that the early-type
  stars ionizing the Gum nebula would excite only a very localized
  region in CG~1. This leaves shock excited \Htwo \ as the only
  viable option. On similar grounds,} MHO 1411, which lies 90\arcsec
\ west of the YSO, is likely to be a highly obscured Herbig-Haro
object. The higher \ceo excitation temperature around the YSO
relative to \ceoNW \ also points to an additional energy source, probably the
YSO.  An indirect indication of shocks is the [HCN]/[HNC] molecular
abundance ratio being found to be larger than 1  by
\citetalias{harjuetal1990}. Such a ratio can be produced by shock
chemistry.

\begin{figure}
\centering
\includegraphics [width=8cm] {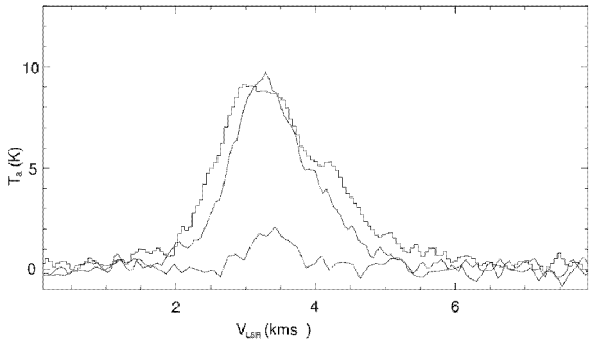}
    \caption{APEX CO (3--2) spectra in the direction of the YSO
      (histogram) and in a $\pm$10'' slice in declination (0,0
      position excluded) (line) and a Hanning smoothed \ceo (3--2)
      spectrum of the centre position. The temperature unit is
      \tastar.  The spectra were obtained from the ESO data archive
      (ESO project 077.C-0217A).  }
\label{fig:archive}
\end{figure}

Even though the typical tracers of an outflow are present (molecular
hydrogen {and Br$\gamma$\ line emission }indicating shocks and a
probable heavily obscured HH object), actual outflow can
be seen neither in the \citetalias{harjuetal1990} CO velocity interval plots
nor the \citet {vanKempenetal2009} CO (3--2) spectrum in the
direction of the YSO. The ESO data archive contains a $\sim$1' by 1\farcm5
on the fly and a smaller pointed CO (3--2) mapping in the direction of
the YSO observed at APEX (ESO project 077.C-0217A, van Kempen et
al.). A regridded, summed-up CO (3--2) map is shown in
Fig. \ref{fig:32map}.  No obvious molecular outflow can be seen in
this data set. The CO (3--2) (histogram) spectrum in the direction of the
YSO is shown in Fig. \ref{fig:archive}. There is an indication of a
very modest redshifted outflow wing in the spectrum. However, part of
this wing may be caused by  a second cloud velocity component seen in the
lower left in Fig. \ref{fig:32map}.  The second spectrum (line plot)
in the figure is a sum of the CO (3--2) spectra in a $\pm$10'' EW
slice in declination through the YSO. The 0,0 position is not included
in the sum. A  Hanning-smoothed \ceo (3--2) 0,0 position spectrum from
the same data set is also shown in the figure.  The morphology of the
YSO nebulosity in the SOFI images and especially the location of MHO
1411 west of the YSO implies that the associated molecular outflow, if
present, should be oriented in the EW direction. However, there is no
indication of outflow wings in the declination-slice summed-up
spectrum. In contrast, it is more narrow than the spectrum in the
centre position.  The \ceo (3--2) spectrum is centred on the CO line.

It is puzzling why no molecular outflow from the YSO is detected.  If
an outflow were present, it would have to be highly collimated and
exactly in the plane of the sky to remain unnoticed.

\section{Summary and conclusions} \label{sec:summary} 

We have combined our new NIR imaging, photometry, spectroscopy, and
{mm} molecular line observations  with already existing data
at other wavelengths to analyse the structure of the globule
head in greater detail than  possible before.  The conclusions are
the following: 

{1. A young stellar object (a likely class II source)
  with an associated bright NIR nebulosity and a molecular hydrogen
  object, MHO1411, were detected in the globule head.  The YSO is
  totally obscured in \Js. The estimated optical extinction towards
  the YSO is 50 magnitudes or more. The \Htwo \ and Br$\gamma$\ line
  emission is detected in the direction of the YSO.  The \Htwo \ emission
  is also seen in the YSO\_SE filament.  }

2. The visual extinction, as estimated with the NICER method, in the
imaged area excluding the YSO direction, ranges from 1\fm3 to 9\fm2
magnitudes.  The maximum extinction lies 40\arcsec \ NW of the YSO.
The peak N(\Htwo) column density estimated from the extinction data is
$9.0 \times 10^{21}cm^{-2}$ and the total mass 16.7\Msun$(d/300pc)^2$. The
mass within the A$_v$= 4~magnitudes contour is 9.2\Msun$ (d/300pc)^2$.

3. The \ceo emission in the globule head is distributed in an
4\arcmin\ by 1\farcm5 area with a sharp maximum {offset SW of the
} YSO. The \ceo \Tmb(2--1) is equal to \Tmb(1--0) in the NW part of
the cloud (the Av maximum) but is significantly stronger than in the
\ceo (1--0) transition in the \ceo maximum and in the SE. This
indicates that the \ceo excitation temperature is higher in the latter
two positions. The \ceo {centre of }line velocity is
3.33$\pm$0.03~\kmps \ in the \ceo maximum and 3.69$\pm$0.03~\kmps \ in
SE indicating, that these are physically separate entities.  It is
likely that the elevated \ceo excitation temperature SW of the YSO is
caused by the interaction of the YSO with the surrounding cloud. The SE
``nose'' coincides with a bright reflection nebulosity south of NX Pup
in the SOFI and SIRIUS images. An  elevated \ceo excitation
temperature in the ``nose'' is likely because of  heating by NX Pup.

4. The peak N(\Htwo) column densities and masses, as calculated from
the \ceo data, are ($5.0 \times 10^{21}cm^{-2}$, 0.75\Msun$~(d/300pc)^2$),
($7.8 \times 10^{21}cm^{-2}$, 9.2\Msun$(d/300pc)^2$) and
($1.35 \times 10^{21}cm^{-2}$, 1.6\Msun$(d/300pc)^2$) for \ceoSE,
\ceomax,  and \ceoNW, respectively. {The column densities and the
  masses are calculated from only two \ceo transitions and should be
considered as very rough estimates}.  The \ceo  masses are lower  than
those calculated from the extinction data.  The likely reason for this
is that the cloud low density envelope is not visible in \ceo.  

{
5. The peak integrated \ceo emission does not coincide with the
position of maximum visual extinction. This is primarily  due to the \ceo
excitation temperature, which is lower in the extinction maximum
than in the \ceo emission maximum.
}

6.  The IRAS point source 07178--4429 resolves into two sources in the
HIRES enhanced IRAS images.  The 12 and 25\,$\mum$\ emission
originates mainly in the Herbig~AeBe star NX Puppis and the 60 and
100$\mum$ emission in the adjacent YSO.{The results of the HIRES analysis are
  confirmed by Spitzer 24$\mum$ and 70$\mum$ MIPS mapping, which has
  become publicly available.} The 60 and 100\,$\mum$\ FIR
luminosity of the point source is 3.1 \Lsun.

7. Even though typical
signs of YSO--parent cloud interaction (shocked gas, elevated \ceo
excitation temperature, MHO object) were detected no strong molecular outflow
seems to be present.

8. If the binary NX Pup and the T Tau star adjacent to it were formed in CG~1,
they have already cleared away the  surrounding dust and molecular gas.
The YSO detected in NIR in the globule head is the second generation of
stars formed in the cloud.

{The new NIR photometry  and molecular line data presented in this paper
  reveal a newly born star and a rather complex distribution of
  interstellar material in the globule head. However, the derived
  cloud parameters, e.g.,  mass, column density, and excitation
  temperature, are only rough estimates.  Further observations are
  necessary to refine the cloud model.  Mapping of the cloud head in
  density sensitive molecules and in both FIR and (sub)mm continuum is needed
  to have sufficient input parameters to run a radiative transfer code like e.g.
the one presented in 
  \citet{juvela1997}. Additional NIR  imaging and spectroscopy
are needed to provide information on the nature of the YSO and the surrounding
nebulosities.}

\begin{acknowledgements}
It is a pleasure to thank the NTT team for the support during the
observing run. This research has made use of the SIMBAD database,
operated at CDS, Strasbourg, France, and of NASA's Astrophysics Data
System Bibliographic Services.This work is based in part on archival
data obtained with the Spitzer Space Telescope, which is operated by
the Jet Propulsion Laboratory, California Institute of Technology
under a contract with NASA. M.M acknowledges the support from the
University of Helsinki Senat's graduate studies grant and from the
Vilho, Yrj\"o and Kalle V\"ais\"al\"a Fund.

\end{acknowledgements}
\bibliographystyle{aa}
\bibliography{14524.bib}

\begin{thebibliography}{37}
\expandafter\ifx\csname natexlab\endcsname\relax\def\natexlab#1{#1}\fi

\bibitem[{{Allen} {et~al.}(1999){Allen}, {Burton}, {Ryder}, {Ashley}, \&
  {Storey}}]{allenetal1999}
{Allen}, L.~E., {Burton}, M.~G., {Ryder}, S.~D., {Ashley}, M.~C.~B., \&
  {Storey}, J.~W.~V. 1999, \mnras, 304, 98

\bibitem[{Ascenso {et~al.}(2007)Ascenso, Alves, Vicente, \&
  Lago}]{ascensoetal2007}
Ascenso, J., Alves, J., Vicente, S., \& Lago, M. 2007, A\&A, 199

\bibitem[{Auman {et~al.}(1990)Auman, Fowler, \& Melnyk}]{aumanetal1990}
Auman, H., Fowler, J., \& Melnyk, M. 1990, AJ, 99, 1674

\bibitem[{{Bernacca} {et~al.}(1993){Bernacca}, {Lattanizi}, {Bucciarelli},
  {Bastian}, {Barabaro}, {Pannunzio}, {Badiali}, {Cardini}, \&
  {Emanuele}}]{Bernaccaetal1993}
{Bernacca}, P.~L., {Lattanizi}, M.~G., {Bucciarelli}, B., {et~al.} 1993, \aap,
  278, L47

\bibitem[{Bertin \& Arnouts(1996)}]{bertinarnouts1996}
Bertin, E. \& Arnouts, S. 1996, A\&AS, 117, 393

\bibitem[{Bessell \& Brett(1988)}]{besselbrett1988}
Bessell, M. \& Brett, J. 1988, PASP, 100, 1134

\bibitem[{{Black} \& {Dalgarno}(1976)}]{blackdalgarno1976}
{Black}, J.~H. \& {Dalgarno}, A. 1976, \apj, 203, 132

\bibitem[{{Blondel} \& {Djie}(2006)}]{blondel2006}
{Blondel}, P.~F.~C. \& {Djie}, H.~R.~E.~T.~A. 2006, \aap, 456, 1045

\bibitem[{{Bohlin} {et~al.}(1978){Bohlin}, {Savage}, \&
  {Drake}}]{bohlinetal1978}
{Bohlin}, R.~C., {Savage}, B.~D., \& {Drake}, J.~F. 1978, \apj, 224, 132

\bibitem[{Bourke {et~al.}(1995)Bourke, Hyland, Robinson, James, \&
  Wright}]{bourkeetal1995a}
Bourke, T., Hyland, A., Robinson, G., James, S., \& Wright, C. 1995, MNRAS,
  276, 1067

\bibitem[{{Brand} {et~al.}(1983){Brand}, {Hawarden}, {Longmore}, {Williams}, \&
  {Caldwell}}]{brandetal1983}
{Brand}, P.~W.~J.~L., {Hawarden}, T.~G., {Longmore}, A.~J., {Williams}, P.~M.,
  \& {Caldwell}, J.~A.~R. 1983, \mnras, 203, 215

\bibitem[{{Brandner} {et~al.}(1995){Brandner}, {Bouvier}, {Grebel}, {Tessier},
  {de Winter}, \& {Beuzit}}]{brandneretal1995}
{Brandner}, W., {Bouvier}, J., {Grebel}, E.~K., {et~al.} 1995, \aap, 298, 818

\bibitem[{{Davis} {et~al.}(2010){Davis}, {Gell}, {Khanzadyan}, {Smith}, \&
  {Jenness}}]{MHO2010}
{Davis}, C.~J., {Gell}, R., {Khanzadyan}, T., {Smith}, M.~D., \& {Jenness}, T.
  2010, \aap, 511, A24+

\bibitem[{{Emerson}(1988)}]{emerson1988}
{Emerson}, J.~P. 1988, in NATO ASIC Proc. 241: Formation and Evolution of Low
  Mass Stars, ed. {A.~K.~Dupree \& M.~T.~V.~T.~Lago}, 21--

\bibitem[{{Foster} {et~al.}(2008){Foster}, {Rom{\'a}n-Z{\'u}{\~n}iga},
  {Goodman}, {Lada}, \& {Alves}}]{hunting2008}
{Foster}, J.~B., {Rom{\'a}n-Z{\'u}{\~n}iga}, C.~G., {Goodman}, A.~A., {Lada},
  E.~A., \& {Alves}, J. 2008, \apj, 674, 831

\bibitem[{{Franco}(1990)}]{franco1990}
{Franco}, G.~A.~P. 1990, \aap, 227, 499

\bibitem[{{Frerking} {et~al.}(1982){Frerking}, {Langer}, \&
  {Wilson}}]{frerkingetal1982}
{Frerking}, M.~A., {Langer}, W.~D., \& {Wilson}, R.~W. 1982, \apj, 262, 590

\bibitem[{Haikala {et~al.}(2006)Haikala, Juvela, Harju, Lehtinen, Mattila, \&
  Dumke}]{haikalaetal2006}
Haikala, L.~K., Juvela, M., Harju, J., {et~al.} 2006, A\&A, 454, L71

\bibitem[{{Haikala} \& {Reipurth}(2010)}]{haikalareipurth2010}
{Haikala}, L.~K. \& {Reipurth}, B. 2010, \aap, 510, A1+

\bibitem[{Harju {et~al.}(1990)Harju, Sahu, C., Wilson, Sahu, \&
  Pottasch}]{harjuetal1990}
Harju, J., Sahu, M., C., H., {et~al.} 1990, A\&A, 233, 197

\bibitem[{Hawarden \& Brand(1976)}]{HawardenBrand1976}
Hawarden, T. \& Brand, P. 1976, MNRAS, 175, 19P

\bibitem[{{Juvela}(1997)}]{juvela1997}
{Juvela}, M. 1997, \aap, 322, 943

\bibitem[{Lada {et~al.}(1994)Lada, Lada, Clemens, \& Bally}]{ladalada1994}
Lada, C., Lada, E., Clemens, D., \& Bally, J. 1994, \apj, 429, 694

\bibitem[{{Lombardi}(2005)}]{lombardi2005}
{Lombardi}, M. 2005, \aap, 438, 169

\bibitem[{Lombardi \& Alves(2001)}]{nicer}
Lombardi, M. \& Alves, J. 2001, A\&A, 377, 1023

\bibitem[{{Malbet}(2007)}]{malbet2007}
{Malbet}, F. 2007, in IAU Symposium, Vol. 243, IAU Symposium, ed. {J.~Bouvier
  \& I.~Appenzeller}, 123--134

\bibitem[{{Nagayama} {et~al.}(2003){Nagayama}, {Nagashima}, {Nakajima},
  {Nagata}, {Sato}, {Nakaya}, {Yamamuro}, {Sugitani}, \&
  {Tamura}}]{Nagayamaetal2003}
{Nagayama}, T., {Nagashima}, C., {Nakajima}, Y., {et~al.} 2003, in Society of
  Photo-Optical Instrumentation Engineers (SPIE) Conference Series, Vol. 4841,
  Society of Photo-Optical Instrumentation Engineers (SPIE) Conference Series,
  ed. {M.~Iye \& A.~F.~M.~Moorwood}, 459--464

\bibitem[{Persson {et~al.}(1998)Persson, Murphy, Krzeminski, Roth, \&
  Rieke}]{perssonetal1998}
Persson, S.~E., Murphy, D.~C., Krzeminski, W., Roth, M., \& Rieke, M.~J. 1998,
  AJ, 116, 2475

\bibitem[{{Puget} \& {Leger}(1989)}]{pugetleger1989}
{Puget}, J.~L. \& {Leger}, A. 1989, \araa, 27, 161

\bibitem[{{Reipurth}(1983)}]{reipurth1983}
{Reipurth}, B. 1983, \aap, 117, 183

\bibitem[{{Sanduleak}(1974)}]{sandauleak1974}
{Sanduleak}, N. 1974, Information Bulletin on Variable Stars, 880, 1

\bibitem[{Santos {et~al.}(1998)Santos, Yun, Santos, \& Marreiros}]{santos1998}
Santos, N., Yun, J., Santos, S., \& Marreiros, R. 1998, A\&A, 116, 1376

\bibitem[{{Smith}(1995)}]{smith1995}
{Smith}, M.~D. 1995, \aap, 296, 789

\bibitem[{{van der Tak} {et~al.}(2007){van der Tak}, {Black}, {Sch{\"o}ier},
  {Jansen}, \& {van Dishoeck}}]{radex2007}
{van der Tak}, F.~F.~S., {Black}, J.~H., {Sch{\"o}ier}, F.~L., {Jansen}, D.~J.,
  \& {van Dishoeck}, E.~F. 2007, \aap, 468, 627

\bibitem[{{van Kempen} {et~al.}(2009){van Kempen}, {van Dishoeck},
  {Hogerheijde}, \& {G{\"u}sten}}]{vanKempenetal2009}
{van Kempen}, T.~A., {van Dishoeck}, E.~F., {Hogerheijde}, M.~R., \&
  {G{\"u}sten}, R. 2009, \aap, 508, 259

\bibitem[{{Warin} {et~al.}(1996){Warin}, {Benayoun}, \&
  {Viala}}]{warinetal1996}
{Warin}, S., {Benayoun}, J.~J., \& {Viala}, Y.~P. 1996, \aap, 308, 535

\bibitem[{{Zinnecker} {et~al.}(1999){Zinnecker}, {Krabbe}, {McCaughrean},
  {Stanke}, {Stecklum}, {Brandner}, {Padgett}, {Stapelfeldt}, \&
  {Yorke}}]{zinneckeretal1999}
{Zinnecker}, H., {Krabbe}, A., {McCaughrean}, M.~J., {et~al.} 1999, \aap, 352,
  L73

\end{thebibliography}
%\listofobjects
\Online
\begin{appendix}
\section{Sofi photometry}
\begin{figure*}[h]
\includegraphics [width=18cm] {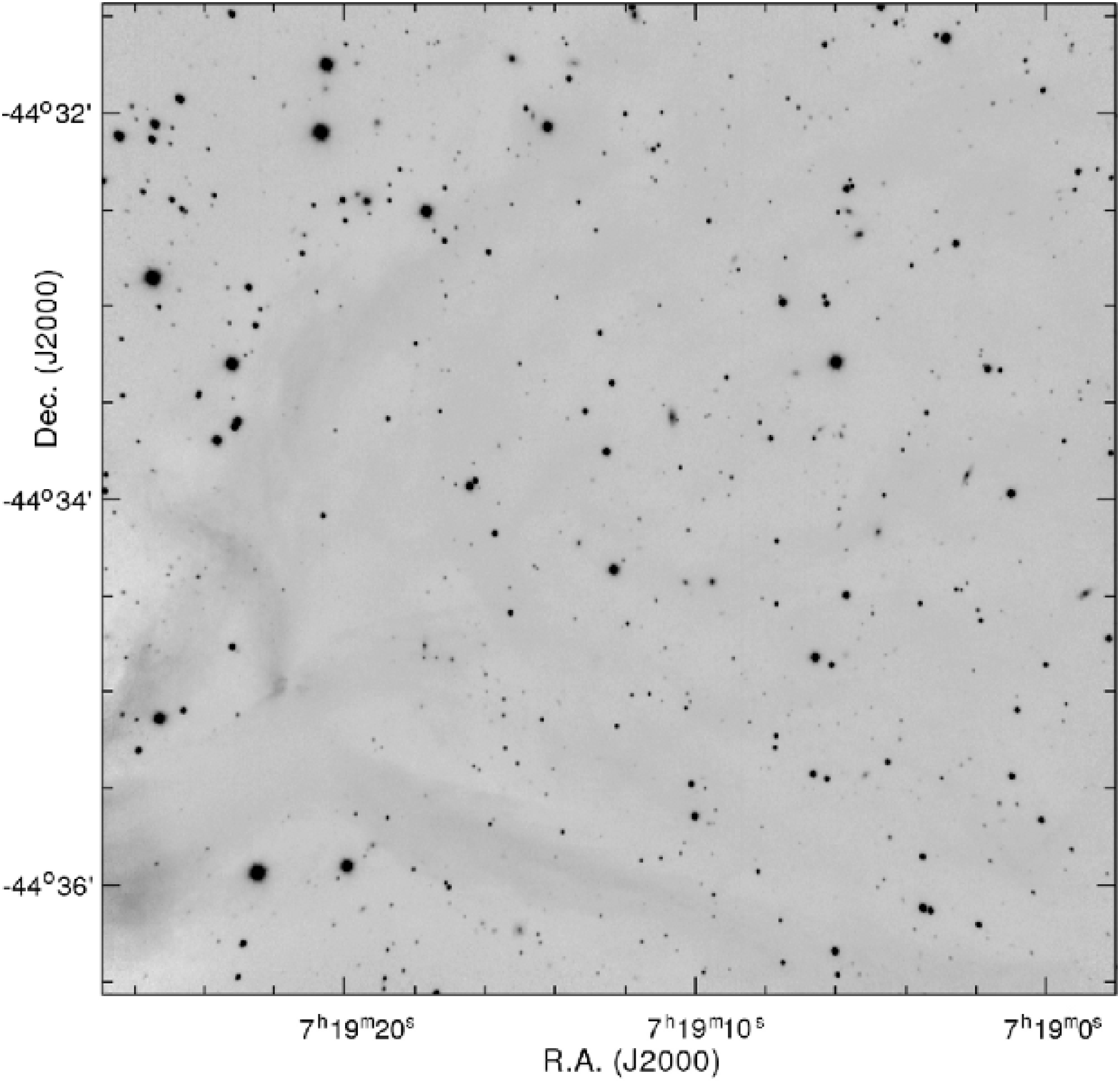}
    \caption{SOFI \Js \ image of CG~1 head.  Square root scaling has
      been used to enhance the appearance of the faint surface
      brightness structures.}
\label{figure:online_Js}
\end{figure*}
\begin{figure*}[h]
\includegraphics [width=18cm] {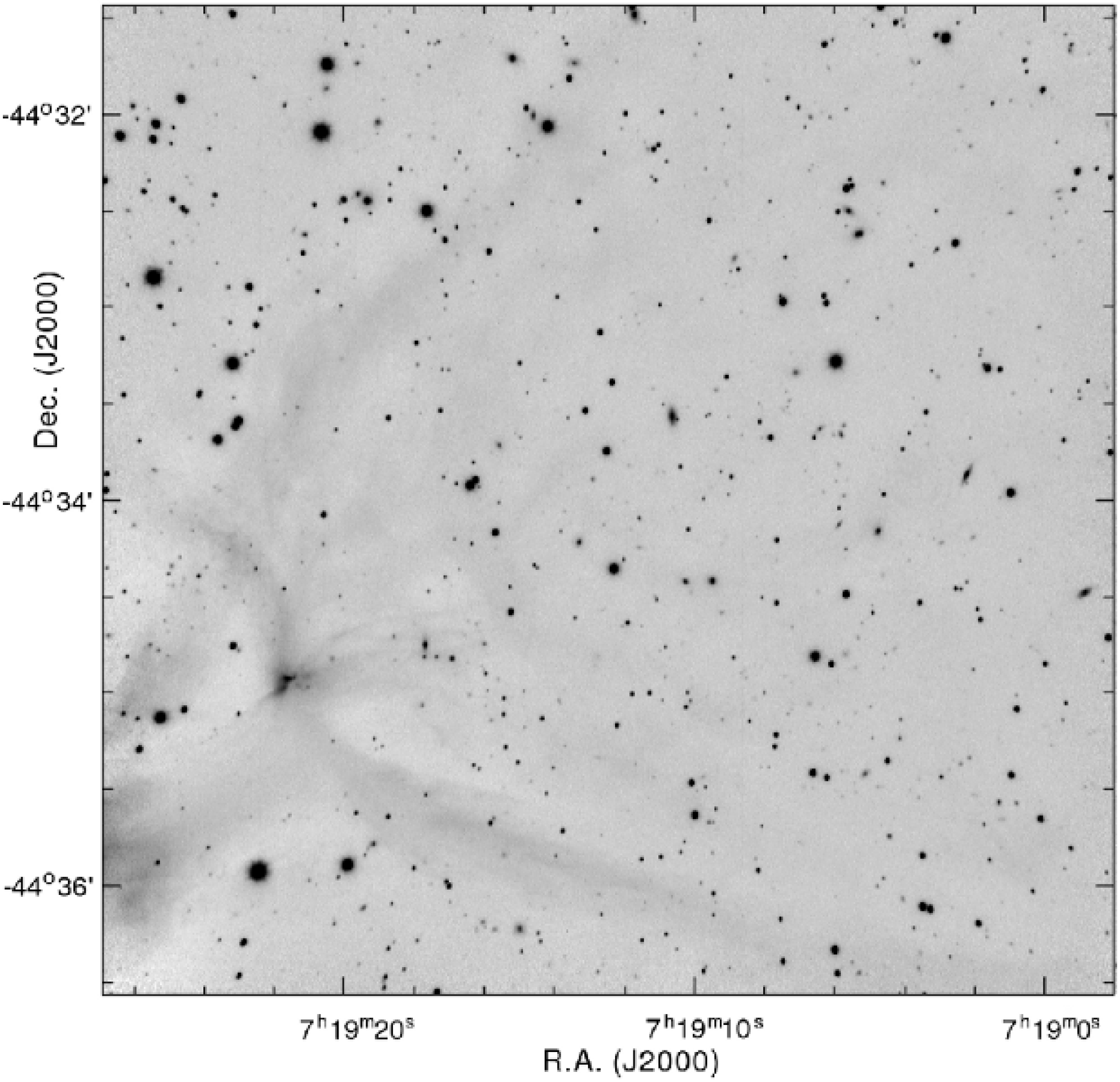}
    \caption{SOFI \H\ image of CG~1 head.  Square root scaling has
      been used to enhance the appearance of the faint surface
      brightness structures.}
\label{figure:online_H}
\end{figure*}
\begin{figure*}[h]
\includegraphics [width=18cm] {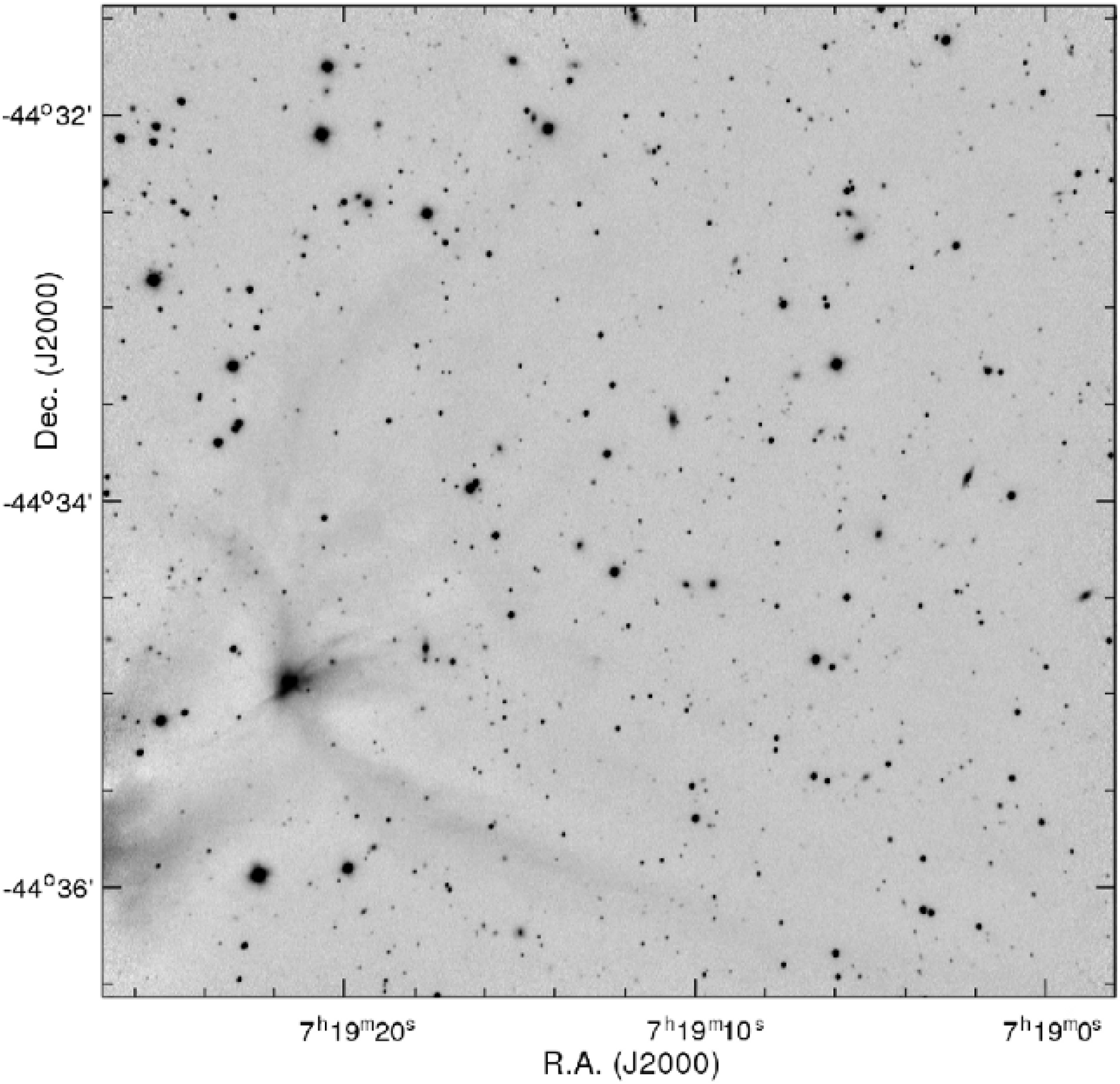}
    \caption{SOFI \Ks \ image of CG~1 head.  Square root scaling has
      been used to enhance the appearance of the faint surface
      brightness structures.}
\label{figure:online_Ks}
\end{figure*}

\begin{figure}[h]
\includegraphics [width=9cm,angle=-90] {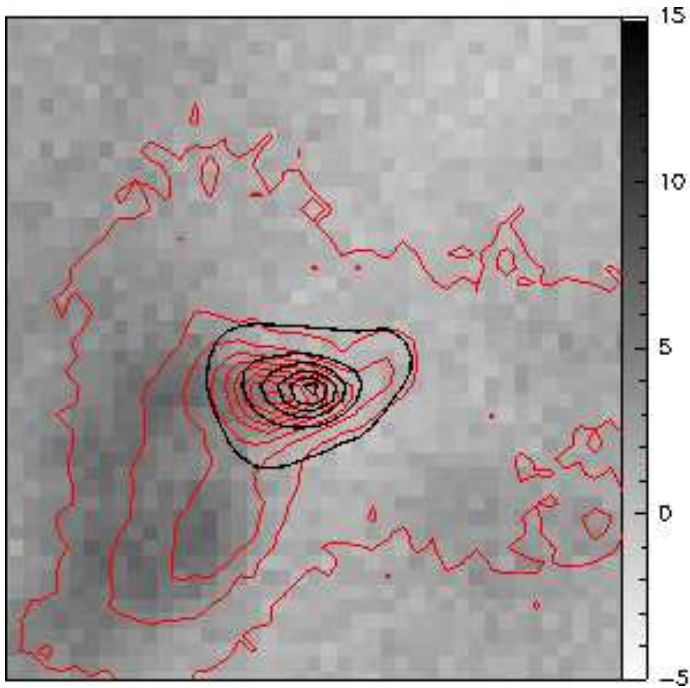}
    \caption{SOFI \Ks \ (in black) and \H \ (in red) surface
      brightness contours superimposed  on a grey scale \Js \ image. The
      contour levels in SOFI counts are from 100 to 1100 in steps of
      100 in \Ks \ and from 10 to 90 in steps of 10 in \H. The wedge
      in the right shows the \Js \ grey scale levels. The image size is 23''
      by 23'' and SOFI pixel scale is 0\farcs288.}
\label{fig:YSOcontours}
\end{figure}

\section{\ceo mapping}
\begin{figure*}[h]
\includegraphics [width=13cm] {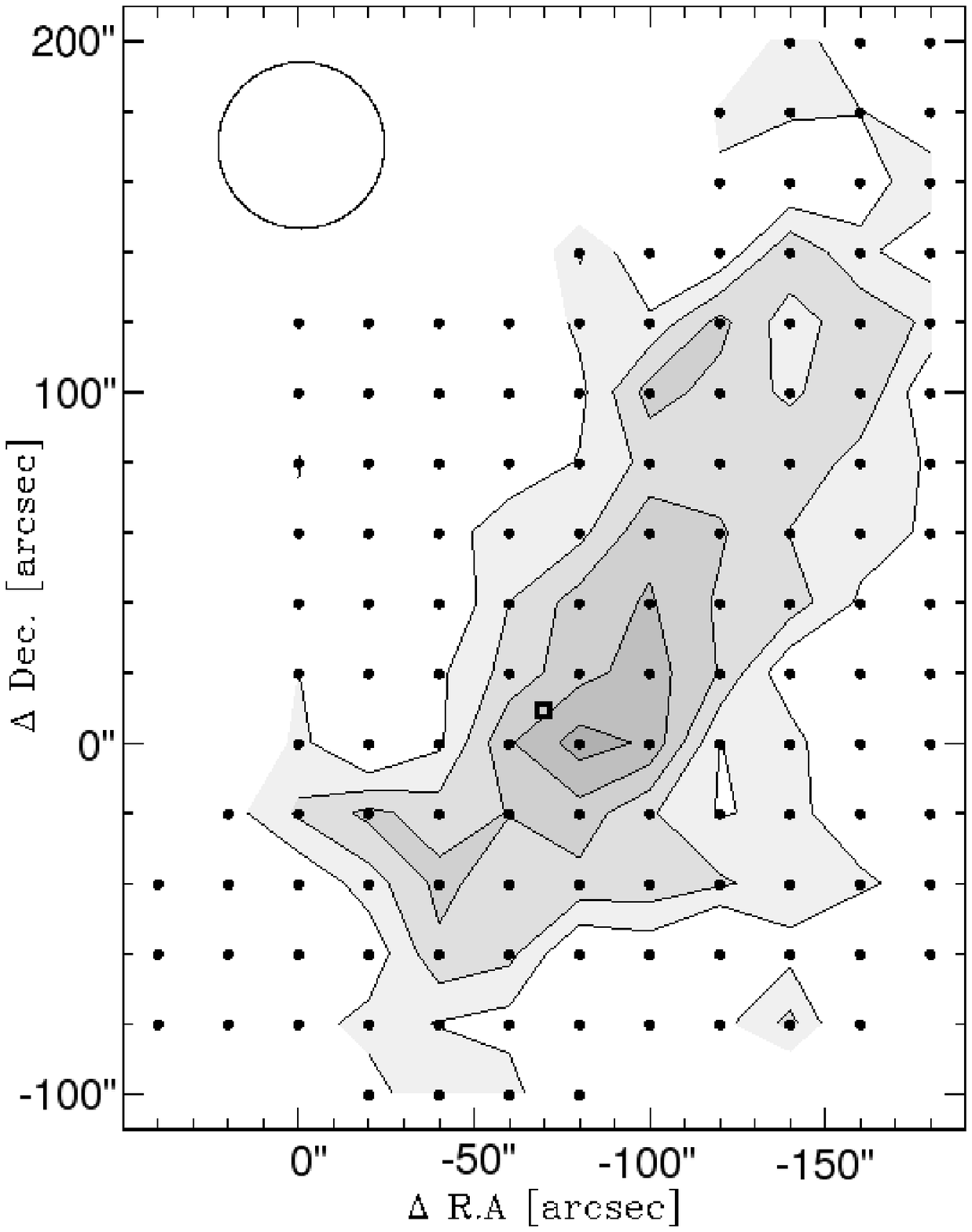}
    \caption{Contour map of the integrated \ceo (1--0) emission in CG~1 head in
      \Tmb \ scale. The contour levels are from 0.25 \kkms \ to 1.25
      \kkms \ in steps of 0.25 \kkms. The points indicate the
      measured points and the square the location of the YSO in the
      figure. The SEST HPBW at the \ceo (1--0) frequency is shown in
      the upper left corner.}
\label{figure:online_1-0}
\end{figure*}
\newpage
\begin{figure*}[h]
\centering
\includegraphics [width=13cm] {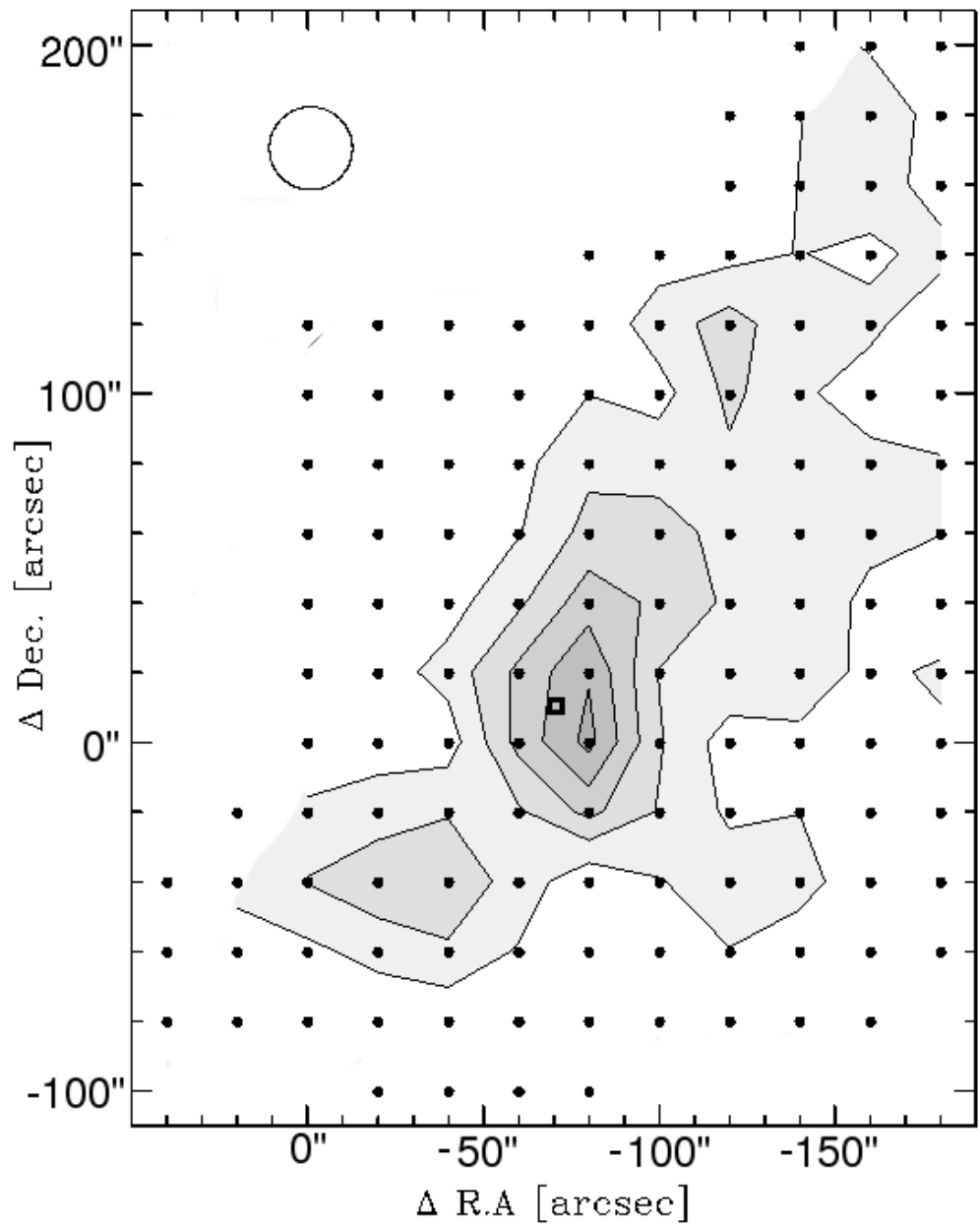}
    \caption{Contour map of the integrated \ceo (2--1) emission in CG~1 head in
      \Tmb \ scale. The contour levels are from 0.4 \kkms \ to 2.0
      \kkms \ in steps of 0.4 \kkms.  The points indicate the
      measured points and the square the location of the YSO in the
      figure. The SEST HPBW at the \ceo (2--1) frequency is shown in
      the upper left corner. }
\label{figure:online_2-1}
\end{figure*}

\section{ESO archive data}
\begin{figure}[h]
\centering
\includegraphics [width=8cm,angle=-90] {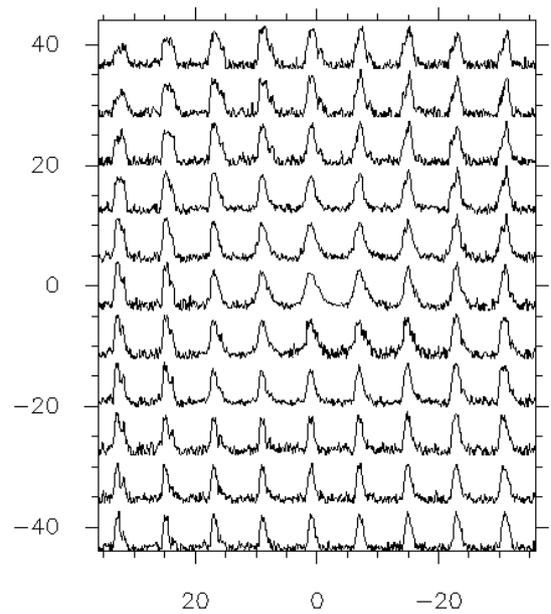}
    \caption{CO (3--2) map in the direction of the YSO. The spectra
      are plotted from 1 \kmps \ to 7\kmps \ in velocity and from -1 K to 12
      K in \tastar \ scale. The offsets are in arcseconds in right
      ascension and declination from the YSO position. The data is ESO
      archive data (ESO project 077.C-0217A)}
\label{fig:32map}
\end{figure}

\end{appendix}
\end{document}